\documentclass[twocolappendix]{emulateapj}
\usepackage{graphicx}
\usepackage{amsmath}

\def\solmass{$\rm M_{\sun}$}
\def\solum{$\rm L_{\sun}$}

\newcommand{\hii}{H{\sc II}}

\begin{document}

\title{PHAT Stellar Cluster Survey I. Year 1 Catalog and Integrated Photometry}

\author{L. Clifton Johnson\altaffilmark{1}, Anil C. Seth\altaffilmark{2}, Julianne J. Dalcanton\altaffilmark{1}, Nelson Caldwell\altaffilmark{3}, Morgan Fouesneau\altaffilmark{1}, Dimitrios A. Gouliermis\altaffilmark{4,5}, Paul W. Hodge\altaffilmark{1}, S{\o}ren S. Larsen\altaffilmark{6}, Knut A. G. Olsen\altaffilmark{7}, Izaskun San Roman\altaffilmark{8}, Ata Sarajedini\altaffilmark{8}, Daniel R. Weisz\altaffilmark{1}, Benjamin F. Williams\altaffilmark{1}, Lori C. Beerman\altaffilmark{1}, Luciana Bianchi\altaffilmark{9}, Andrew E. Dolphin\altaffilmark{10}, L\'eo Girardi\altaffilmark{11}, Puragra Guhathakurta\altaffilmark{12}, Jason Kalirai\altaffilmark{13}, Dustin Lang\altaffilmark{14}, Antonela Monachesi\altaffilmark{15}, Sanjay Nanda\altaffilmark{16}, Hans-Walter Rix\altaffilmark{5}, and Evan D. Skillman\altaffilmark{17}}

\email{lcjohnso@astro.washington.edu}
\altaffiltext{1}{Department of Astronomy, University of Washington, Box 351580, Seattle, WA 98195, USA}
\altaffiltext{2}{Department of Physics and Astronomy, University of Utah, Salt Lake City, UT 84112, USA}
\altaffiltext{3}{Harvard-Smithsonian Center for Astrophysics, 60 Garden Street, Cambridge, MA 02138, USA}
\altaffiltext{4}{Institut f\"ur Theoretische Astrophysik, Zentrum f\"ur Astronomie der Universit\"at Heidelberg, Albert-Ueberle-Stra{\ss}e~2, 69120 Heidelberg, Germany}
\altaffiltext{5}{Max-Planck-Institut f\"ur Astronomie, K\"onigstuhl 17, 69117 Heidelberg, Germany}
\altaffiltext{6}{Department of Astrophysics, IMAPP, Radboud University Nijmegen, P.O. Box 9010, 6500 GL Nijmegen, The Netherlands}
\altaffiltext{7}{National Optical Astronomy Observatory, 950 North Cherry Avenue, Tucson, AZ 85719, USA}
\altaffiltext{8}{Department of Astronomy, University of Florida, 211 Bryant Space Science Center, Gainesville, FL 32611-2055, USA}
\altaffiltext{9}{Department of Physics and Astronomy, Johns Hopkins University, 3400 North Charles Street, Baltimore, MD 21218, USA}
\altaffiltext{10}{Raytheon Company, 1151 East Hermans Road, Tucson, AZ 85756, USA}
\altaffiltext{11}{Osservatorio Astronomico di Padova -- INAF, Vicolo dell'Osservatorio 5, I-35122 Padova, Italy}
\altaffiltext{12}{University of California Observatories/Lick Observatory, University of California, 1156 High Street, Santa Cruz, CA 95064, USA}
\altaffiltext{13}{Space Telescope Science Institute, 3700 San Martin Drive, Baltimore, MD 21218, USA}
\altaffiltext{14}{Department of Astrophysical Sciences, Princeton University, Princeton, NJ 08544, USA}
\altaffiltext{15}{Department of Astronomy, University of Michigan, 500 Church Street, Ann Arbor, MI 48109, USA}
\altaffiltext{16}{Indian Institute of Technology, Kanpur, Uttar Pradesh 208016, India}
\altaffiltext{17}{Department of Astronomy, University of Minnesota, 116 Church Street SE, Minneapolis, MN 55455, USA}


\begin{abstract}
The Panchromatic Hubble Andromeda Treasury (PHAT) survey is an on-going Hubble Space Telescope (HST) multi-cycle program to obtain high spatial resolution imaging of one-third of the M31 disk at ultraviolet through near-infrared wavelengths.  In this paper, we present the first installment of the PHAT stellar cluster catalog.  When completed, the PHAT cluster catalog will be among the largest and most comprehensive surveys of resolved star clusters in any galaxy.  The exquisite spatial resolution achieved with HST has allowed us to identify hundreds of new clusters that were previously inaccessible with existing ground-based surveys.  We identify 601 clusters in the Year 1 sample, representing more than a factor of four increase over previous catalogs within the current survey area (390 $\text{arcmin}^2$).  This work presents results derived from the first $\sim$25\% of the survey data; we estimate that the final sample will include $\sim$2500 clusters.  For the Year 1 objects, we present a catalog with positions, radii, and six-band integrated photometry.  Along with a general characterization of the cluster luminosities and colors, we discuss the cluster luminosity function, the cluster size distributions, and highlight a number of individually interesting clusters found in the Year 1 search.
\end{abstract}

\keywords{catalogs --- galaxies: individual (M31) --- galaxies: star clusters: general}

\newpage
\section{Introduction} \label{intro}

Large, high-quality samples of stellar clusters provide key data for studies of a wide variety of astrophysical topics, including cluster evolution, stellar evolution, star formation, and galaxy evolution.  However, current cluster samples suffer from serious limitations.  For example, Milky Way clusters suffer from severe dust attenuation within the Galactic plane, resulting in disk cluster samples that are complete only within a small region (radius of $\sim$1~kpc) around the Sun \citep[e.g.,][]{Dias02, Piskunov08}.  This limits the variety of objects and galactic environments explored, as exemplified by the dearth of massive, intermediate-age clusters known within the Milky Way \citep[e.g.,][]{Davies11}.  Although infrared surveys of the Galactic plane are improving this situation \citep[e.g.,][]{Dutra03, Mercer05, Borissova11}, current samples of Galactic clusters do not probe the full stellar cluster parameter space, limiting our ability to study mass, age, and environmental dependencies of evolutionary processes. 

One solution to the incompleteness of Galactic samples has been to extend cluster studies to other galaxies.  Extragalactic cluster samples have grown immensely over the past decade, thanks in part to the power of Hubble Space Telescope (HST) imaging.  These analyses probe a variety of galactic environments, ranging from starbursting galaxy mergers \citep[e.g.,][]{Whitmore99} to quiescent spirals \citep[e.g.,][]{Larsen02}, producing provocative results concerning the environmental dependence of cluster formation and evolution \citep[e.g.,][]{Larsen00, Goddard10}.  However, even with HST, clusters in most distant galaxies appear as marginally resolved single objects.  This increases the difficulty and uncertainty associated with basic steps in cluster analysis, such as object identification, photometry, and the derivation of ages and masses.  As a result, the interpretation of underlying cluster evolutionary processes has considerable associated uncertainty, even leading to cases of conflicting interpretations derived from the same dataset \citep[e.g., in M83;][]{Chandar10-M83, Bastian12, Fouesneau12}.

Closer to the Milky Way, large cluster catalogs exist for the Large and Small Magellanic Clouds \citep[LMC \& SMC;][]{Bica08, Bica99, Bica00, Hunter03}.  In the Clouds, the ability to resolve clusters into individual stars has resulted in a number of important results in stellar evolution \citep[e.g.,][]{Chiosi89, Frogel90, Girardi09}, cluster evolution \citep[e.g.,][]{Gieles08, Chandar10}, and cluster formation \citep[e.g.,][]{Mackey08}.  However, there are limitations associated even with this excellent sample of objects.  On a practical level, while these cluster samples do not suffer from the same incompleteness issues as those from the Milky Way, their piecemeal assembly by many different groups has imprinted a complex and little-understood selection function.  On a more fundamental level, there are additional limits to the applicability of stellar and cluster evolution results derived from these interacting, relatively low-mass galaxies.  The LMC and SMC do not probe the range in galactic environments that are characteristic of a majority of the baryonic universe.  More than 75\% of all stars in the Universe today have metallicities within a factor of two of the solar value \citep{Gallazzi08}, higher than those probed by the Clouds.  No cluster sample comparable to those in the Clouds is currently available in a large spiral galaxy.

The neighboring galaxy M31 (the Andromeda galaxy) is a prime target for stellar cluster studies.  The galaxy's proximity allows resolution of individual bright stars in clusters and the robust detection of clusters down to faint ($<10^{4}$ \solum) luminosities.  M31 also provides access to a range of widely-varying environments across the extent of the star-forming disk.  Andromeda's role as a valuable laboratory was realized long ago, and decades of work have gone into exploring its cluster population.  We defer the detailed review of existing cluster catalogs until Section \ref{oldcat}, but beginning with \citet{Hubble32}, previous work has mainly focused on M31's globular cluster population through ground-based imaging \citep[e.g.,][]{Galleti04, Peacock10}.  The proximity of M31 also enables high-resolution spectroscopic follow-up of bright clusters \citep[e.g.,][]{Colucci09,Strader11}, as well as low-resolution spectroscopy of intrinsically faint clusters that are inaccessible in distant galaxies \citep[e.g.,][]{Caldwell09}.  Numerous studies have utilized HST's excellent spatial resolution to study massive clusters \citep[e.g.,][]{Barmby01, Perina09} and their individual resolved stars \citep[e.g.,][]{Rich05, Mackey06, Perina09a, Perina11}, as well as to identify and study less massive disk clusters \citep[e.g., the Hodge-Krienke Catalogs;][hereafter the HKC]{Krienke07, Krienke08, Hodge09, Hodge10}.  While extremely valuable, these previous space-based observing programs have been limited to focused studies of a small number of targeted regions, as opposed to a wide-ranging survey to obtain a broad sampling of the galaxy and an overall assessment of the M31 cluster population.

\begin{figure*}[ht!]
\centering
\includegraphics[scale=0.4]{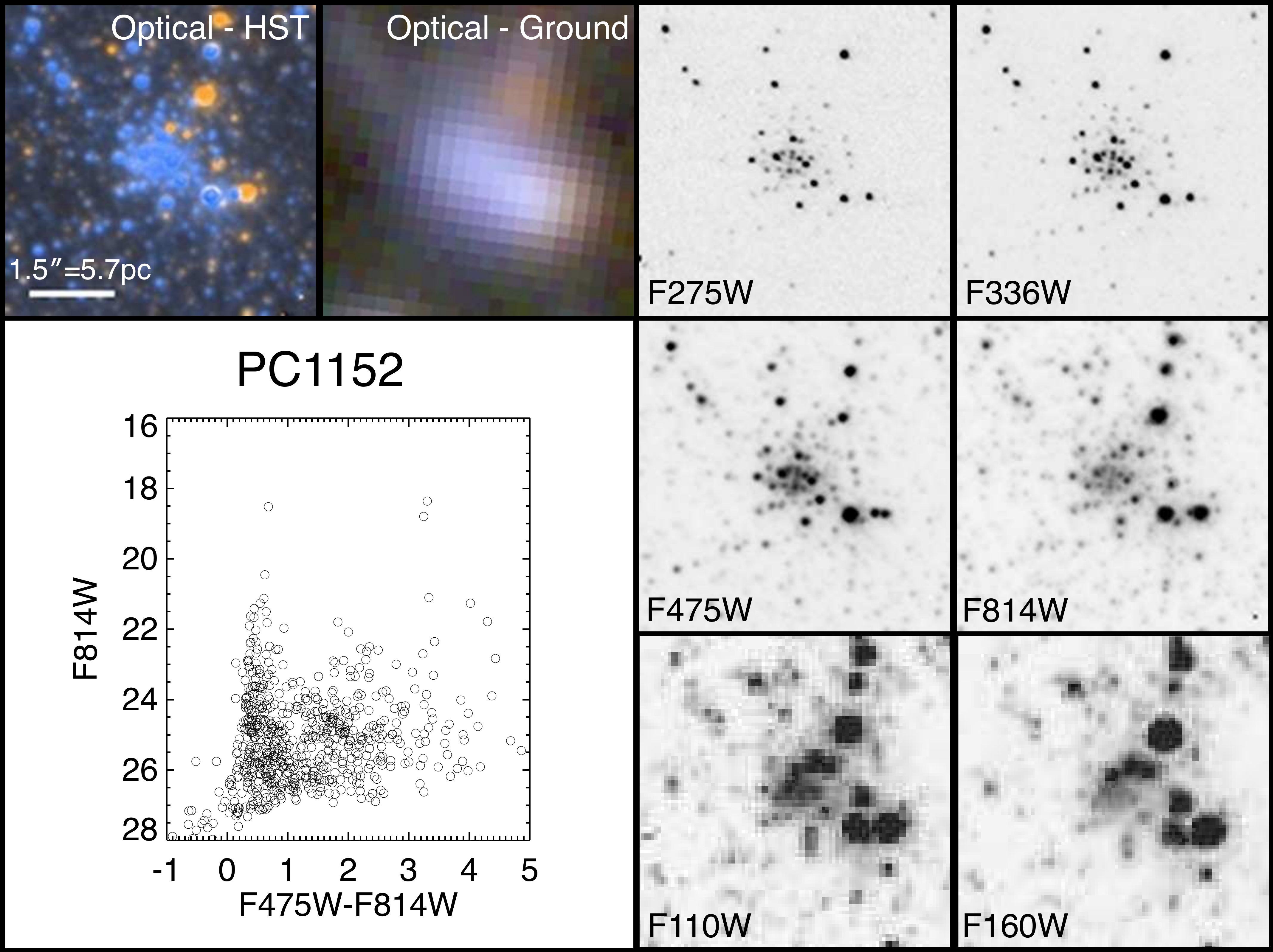}
\caption{PHAT survey data quality example, showing newly-identified cluster PC1152.  Six grayscale single-band images, as well as a color optical (F475W+F814W) mosaic, show the superior image quality provided by HST when compared to ground-based observations \citep[from the Local Group Galaxy Survey;][]{Massey06}.  We also present a color-magnitude diagram of resolved photometry for objects that lie within the cluster cutout image.  The cluster's main sequence forms the vertical sequence at F475W-F814W$\sim$0.5, along with three evolved giants at F814W$<$20.  Other stars shown with F814W$>$20 and F475W-F814W$>$1.0 are likely coincident stars belonging to the background field.}
\label{clstex}
\end{figure*}

The Panchromatic Hubble Andromeda Treasury \citep[PHAT;][]{Dalcanton12} is poised to revolutionize the study of stellar clusters in M31.  This on-going HST multi-cycle program will image one-third of the M31 disk at high spatial resolution, with wavelength coverage from the ultraviolet through the near-infrared.  In terms of cluster studies, this survey provides a number of distinct advantages over existing work.  High spatial resolution imaging allows us to resolve clusters into individual stars, permitting detailed characterization of their stellar populations through analysis of their color-magnitude diagrams (CMDs).  Figure~\ref{clstex} demonstrates the data quality provided by PHAT for a previously unidentified cluster, showing the considerable gain over existing ground-based surveys.  The wide wavelength coverage enables accurate age-dating, as well as the ability to probe a broad range of stellar effective temperatures, from massive main sequence stars to evolved supergiants and AGB stars.  As a result, these objects provide a wealth of valuable knowledge in terms of calibrating and refining stellar evolution models at high metallicity.  Finally, these images allow the detection of faint clusters, reliably probing more than $\sim$2 magnitudes further down the luminosity function than was previously possible using ground-based datasets.  The PHAT survey will increase the HST spatial coverage of the M31 searched for clusters by a factor of $\sim$3 (1800 $\text{arcmin}^2$ in total compared to 650 $\text{arcmin}^2$ surveyed in the HKC; 390 $\text{arcmin}^2$ in the current Year 1 dataset).  However, this metric underestimates the scientific gain provided by high-quality, uniform PHAT data products, as opposed to the heterogeneous archival data used in the HKC work.  Fundamentally, the PHAT survey represents the shift from a discrete, targeted mode of cluster study to a broad survey mode, allowing for comprehensive analysis of cluster evolutionary processes its environmental dependencies.

The stellar clusters identified as part of the PHAT survey will constitute the most comprehensive sample of clusters available for any large spiral galaxy.  The large range of galactocentric radius (0-20~kpc) included in the survey spans a wide range of star formation intensities and gas densities.  The diversity of galactic environments will be important for testing models of cluster formation and evolution.  In addition, the simultaneous accessibility of objects over a $>$3 order of magnitude range in cluster luminosity provides a top-to-bottom view of the cluster population, given that we sample a continuous range of objects that extend from those equivalent to Galactic open clusters up to massive globular clusters.

This paper is the first in a series utilizing the PHAT dataset for studies of stellar clusters.  Here, we present the first installment of a HST-based cluster catalog, which will serve as the basis for extensive study of Andromeda's cluster population. Catalog updates and improvements will be published over the course of this four year observing program.  In this edition, we publish positions, sizes, and integrated photometry for the Year 1 cluster sample.  Age and mass determinations derived from the integrated photometry will be presented in Fouesneau et al. (2012, in prep.).  Additional studies, including analysis of structural parameters, resolved star content, and integrated spectroscopy of the cluster sample will follow in subsequent work.

We summarize the PHAT observations in Section \ref{data}, while in Section \ref{analysis} we describe our cluster identification procedures, present results from completeness testing, and introduce the Year 1 cluster catalog.  Next, we describe and test our photometry methodology in Section \ref{phot}, followed by a comparison between the PHAT cluster catalog and existing catalogs in Section \ref{oldcat}.  We present a basic characterization of the cluster catalog contents in Section \ref{results}, followed by discussions of luminosity functions, the cluster size distribution, and objects of interest in Section \ref{discuss}.  We conclude with a summary and description of future work in Section \ref{conclusion}.  Throughout this work, we assume a distance modulus for M31 of 24.47 \citep[785 kpc;][]{McConnachie05}, for which $1''$ corresponds to a physical size of 3.81~pc.  

\section{Observational Data} \label{data}

This paper includes clusters identified in Year 1 PHAT imaging data, taken before June 2011.  A full description of the PHAT observational design is available in \citet{Dalcanton12}, but we briefly summarize relevant details below.  PHAT observations are grouped into 23 area units known as ``bricks'', each made up of 18 mosaiced HST fields of view in a rectangular 6$\times$3 arrangement that covers a $\sim$12$'$$\times$6.5$'$ region of sky.  Data are obtained simultaneously at different field centers with the ACS (using the F475W and F814W filters) and WFC3 (using the F275W, F336W, F110W, and F160W filters) instruments in two epochs, separated by $\sim$6 months.  During each epoch, imaging is obtained by the cameras in two side-by-side, half-brick (3$\times$3) arrays.  Between the epochs, the orientation of the cameras change by 180 degrees due to the annual roll angle variation of HST.  As a result, the half-brick mosaic obtained by the ACS camera during the first epoch is now imaged by the WFC3 camera in the second epoch, and vice-versa, completing six-filter imaging across the brick.  In all, we obtain $\sim$130 minutes of exposure time at each of the brick's 18 field centers.

\begin{figure*}
\centering
\includegraphics[scale=1.15]{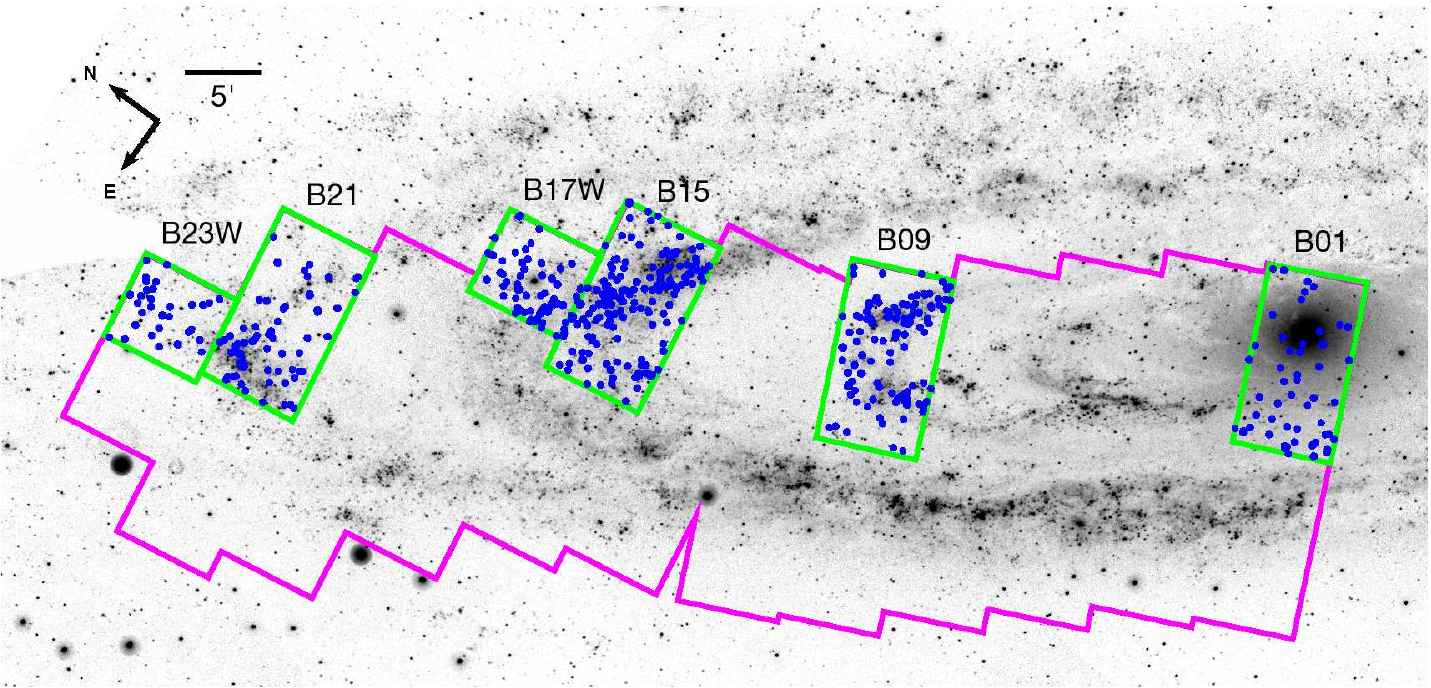}
\caption{Footprint of the PHAT survey region (magenta) displayed on a GALEX NUV image of the northeast half of M31.  Green rectangles represent the ``bricks'' that make up the Year 1 imaging data.  Blue circles show the spatial distribution of clusters identified in the Year 1 cluster search.}
\label{footprint}
\end{figure*}

The Year 1 imaging used in this work includes four full bricks (designated B01, B09, B15, and B21) and the western halves of two additional bricks (B17W and B23W).  These data sample locations along the major axis of M31 from the center out to a projected radius of $\sim$20~kpc. The full PHAT survey footprint, along with the locations of the Year 1 bricks are presented in Figure \ref{footprint}.

Combining imaging data across multiple pointings from three separate cameras (ACS/WFC, WFC3/UVIS, and WFC3/IR) requires astrometry with higher precision than that obtained using the standard telescope telemetry and data processing pipeline.  Images of neighboring fields have sufficient overlap to enable us to derive an astrometric solution across a full (or half) brick.  These astrometric solutions are obtained separately for each camera using photometric catalogs derived using DOLPHOT\footnote{\url{http://purcell.as.arizona.edu/dolphot}}, a modified version of HSTPhot \citep{Dolphin00} that has been updated to include specialized ACS and WFC3 modules.  Affine distortion corrections (in addition to those already known for each camera from the IDCTABs) are required to obtain consistent brick-wide astrometric solutions.  An additional correction brings the brick-wide astrometric solutions onto a global astrometric frame, defined using CFHT observations tied to 2MASS catalogs \citep{Skrutskie06}. The global astrometric alignment agrees with that of the 2MASS reference system within an absolute level of $\sim$60 mas.

Once aligned, we use the \texttt{multidrizzle} task within PyRAF \citep{Koekemoer02} for cosmic ray rejection and image creation along with \texttt{lacosmic} \citep{vanDokkum01} for supplemental cosmic ray flagging.  We note that cosmic ray correction is particularly difficult in the case of the WFC3/UVIS data, due to a large number of cosmic ray artifacts (particularly in F275W images) and the availability of only two frames for artifact detection in regions of non-overlapping field coverage.  Pixel scales of the resulting images are (0.04, 0.05, and 0.065 arcsec/pixel) for the (WFC3/UVIS, ACS/WFC, WFC3/IR) cameras, where the WFC3/IR images are up-sampled from their native plate scale (0.128 arcsec/pixel) to take advantage of the higher effective resolution afforded by the survey's sub-pixel dither strategy.

\section{Cluster Identification} \label{analysis}

The goal of the PHAT cluster survey is to identify and analyze a sample of gravitationally bound stars clusters in M31.  In this paper, we undertake the first step toward achieving this goal: the visual inspection of Year 1 PHAT imaging to identify candidate bound clusters.  As shown in Fig.~\ref{clstex}, clusters appear in PHAT imaging as composite objects composed of a centrally concentrated overdensity of individual resolved stars and a broad unresolved light component.  The ability to resolve these objects into individual stars allows for the clean separation between genuine stellar clusters and contaminants such as background galaxies or single stars.

Although PHAT imaging facilitates robust identifications of stellar overdensities, determining whether or not these clusterings are gravitationally bound is challenging.  Quantitative assessment of an object's boundedness requires age, mass, and spatial profile information \citep[e.g.,][]{Gieles11}.  Further, determining the boundedness of young objects ($\lesssim$10 Myr) is made even more difficult, because dynamical evolution has had little time to evolve stellar structures from an initial hierarchical, scale-free spatial distribution \citep[for further discussion, see e.g.,][and references therein]{Bastian11}, blurring the distinction between bound and unbound stellar groupings.  In future work, we will utilize age, mass, and structural characteristics to assess boundedness for each object (e.g., using the $\Pi$ statistic; \citealt{Gieles11}), but that analysis is beyond the scope of the initial classification work presented here.

Given the current limitations in assessing an object's boundedness, we adopt a liberal approach for cluster candidate identification.  We prioritize sample completeness over purity and therefore include all cluster-like stellar overdensities as part of the object catalog.  We acknowledge that the cluster catalog presented here will likely include both bound and unbound groupings of stars, particularly among the youngest objects.  As a result, throughout this work all objects are
formally considered cluster candidates, although for brevity we will refer to them simply as clusters.

With the intent of the cluster search well defined, we proceed to a description of our search methodology and a presentation of the cluster search results in Section~\ref{byeye}.  We conclude our discussion of cluster identification by characterizing the completeness characteristics of the sample in Section~\ref{comp}.

\begin{figure*}[Ht!]
\centering
\includegraphics[scale=0.6]{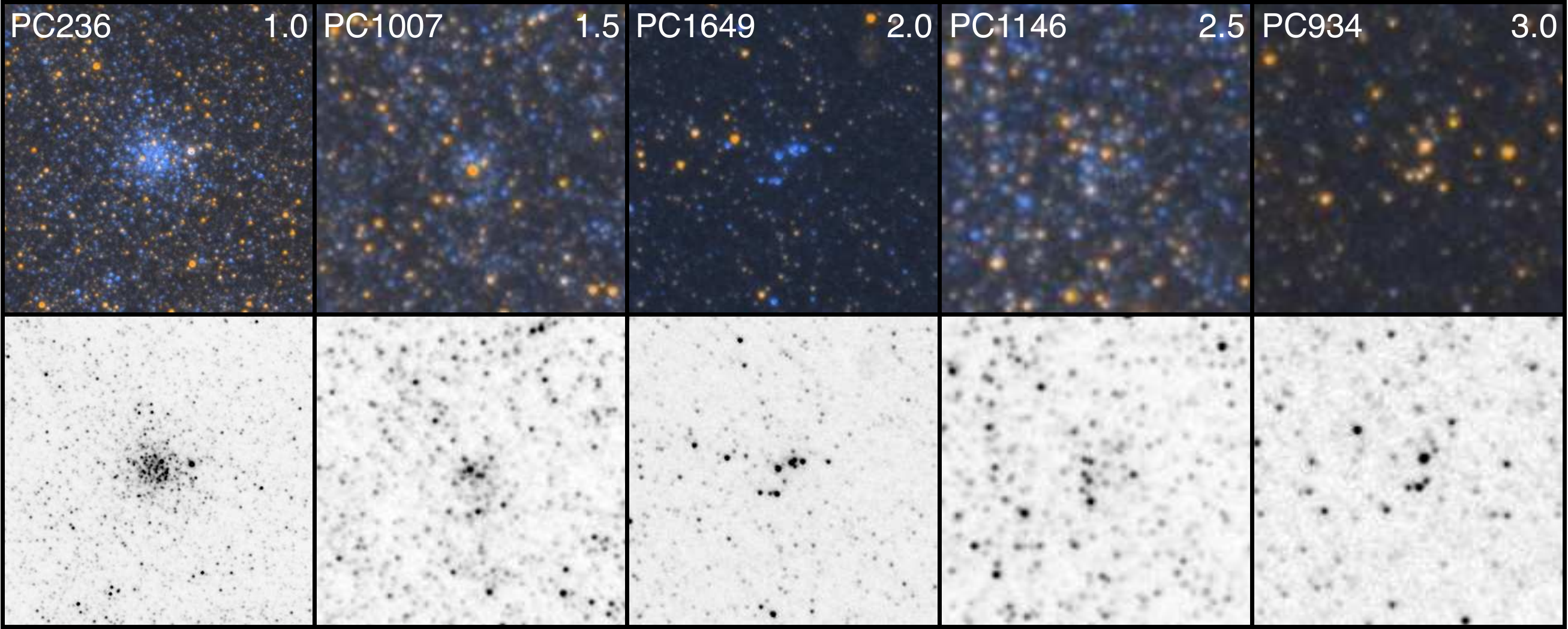}
\caption{Cluster cutout images showing objects across the range of possible $S_{by-eye}$.  The top row shows color optical (F475W+F814W) images, and the bottom row shows grayscale F475W images.  Labels present the object name along with the average $S_{by-eye}$, which runs on a scale of 1 (definite cluster) to 3 (unlikely cluster).}
\label{scoreex}
\end{figure*}

\subsection{By-Eye Search} \label{byeye}

We undertake a systematic by-eye search of the Year 1 PHAT images following a precedent set by previous M31 cluster studies \citep[e.g.,][]{Barmby01,Krienke07}.  Drawing on this rich history of visual cluster identification, we make a concerted effort to improve upon previous work through the use of uniform analysis techniques, redundancy, cross-validation, and improved characterization of the resulting selection function.

When using any method of cluster identification, manual or automated, it is important to understand the biases and limitations inherent to the technique.  We use artificial cluster tests in order to assess the completeness characteristics of our search methodology, which we discuss in detail in Section~\ref{comp}.  We are also developing automated methods of cluster identification to further reduce subjectivity of the identifications and enable more robust completeness testing in future work with this dataset (Olsen et al. 2012, in prep.).  The cluster sample presented here will act as an important comparison sample to help refine these automated search techniques for use with the PHAT dataset.

Our by-eye search consists of two stages: an initial search of the available imaging for all viable cluster candidates, followed by a re-evaluation of each preliminary candidate in a systematic manner. The completeness and accuracy of this process are enhanced by the redundancy of eight experienced astronomers conducting each stage of the search.

To perform the initial image search, the survey area is subdivided using the footprint of the WFC3-IR camera (2.3$'$$\times$2.1$'$), providing 18 contiguous, minimally-overlapping search fields within each brick.  Each field is searched by three or four team members in a ``blind'' manner, meaning that these individuals are not provided with the locations of clusters previously identified by other PHAT team members or previous surveys.  To identify cluster candidates, searchers use a suite of images that include a two-band optical color image, all six available single-band images spanning from the UV to the NIR, and two star-subtracted optical images used to identify diffuse emission that makes up a cluster's unresolved light.  These images are accessed using a custom image viewer that allows users to switch between the available spatially-aligned, full resolution images, as well as the ability to alter brightness and contrast levels for optimal visualization.

Once candidate objects are selected in the initial visual search, we cross-match and combine the identifications of all team members.  Based on this preliminary catalog, we perform initial photometry and create cutout images for each cluster candidate.  The preliminary catalog is reviewed independently by all eight team members on an interactive web site, and each individual assigns scores for each candidate, providing an assessment of the likelihood that an object is a cluster.  Scores ($S_{by-eye}$) are based on a scale of 1-to-3, where $S_{by-eye}=1$ represents a definite cluster, $S_{by-eye}=2$ signifies a likely cluster, and $S_{by-eye}=3$ represents an unlikely or non-cluster object.  We present examples of the cluster scoring system in Fig.~\ref{scoreex}.

We average the scores from all team members for each object and use these average rankings to divide the candidates into three subsamples: clusters, possible clusters, and unlikely objects.  We choose thresholds of $S_{by-eye} < 2.0$ for clusters, $2.0 \leq S_{by-eye} < 2.5$ for possible clusters, and $S_{by-eye} \geq 2.5$ for unlikely objects.  We discard unlikely objects from the catalog, while retaining clusters and possible clusters in two separate catalogs.  The average scores for each cluster are provided as part of the cluster catalogs as an assessment of candidate quality.  To determine the reliability of these average scores, we built a ranking experiment into our classification procedure.  During the course of the ranking work, 24 objects appeared twice within the preliminary catalog.  As a result, these clusters were each ranked two separate times by each team member.  When the resulting average scores of the duplicate entries are compared, we find the standard deviation of the 1-to-3 ranking differences to be 0.27, showing good consistency and repeatability for the scores provided by our search team.

\begin{figure}[ht!]
\centering
\includegraphics[scale=0.4]{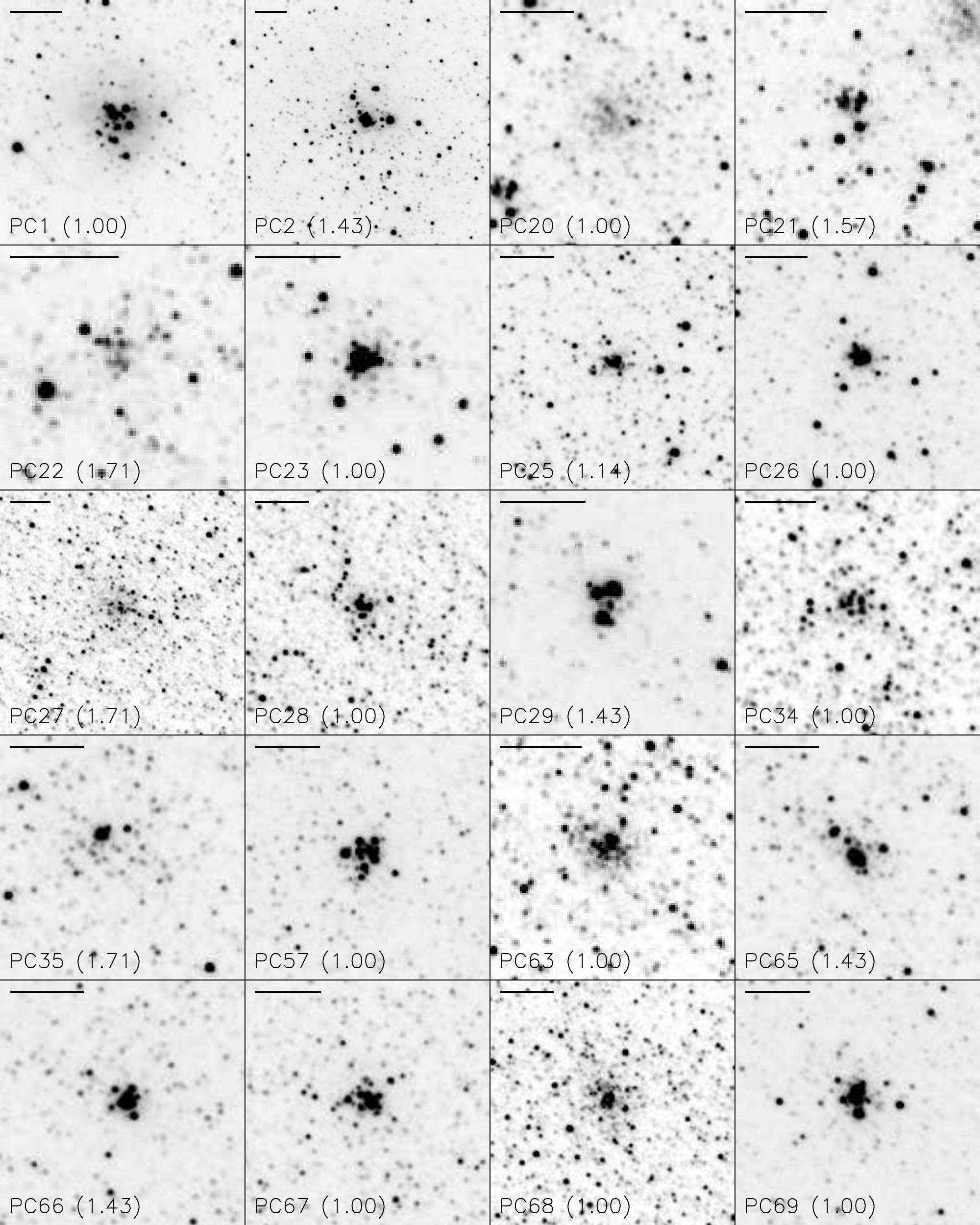}
\caption{Cutout images for clusters in Table \ref{tbl1} with $S_{by-eye} < 2.0$.  Images are grayscale F475W images, scaled to three times the cluster radius, and aligned such that North is up and East is left.  Along with the PHAT cluster identifier, the average $S_{by-eye}$ is provided for each object in parenthesis.  The scale bar in each image represents 2$''$.  Figures \ref{cutout}.1--\ref{cutout}.31 are available in the online version of the Journal.}
\label{cutout}
\end{figure}

\begin{figure}[ht!]
\centering
\includegraphics[scale=0.4]{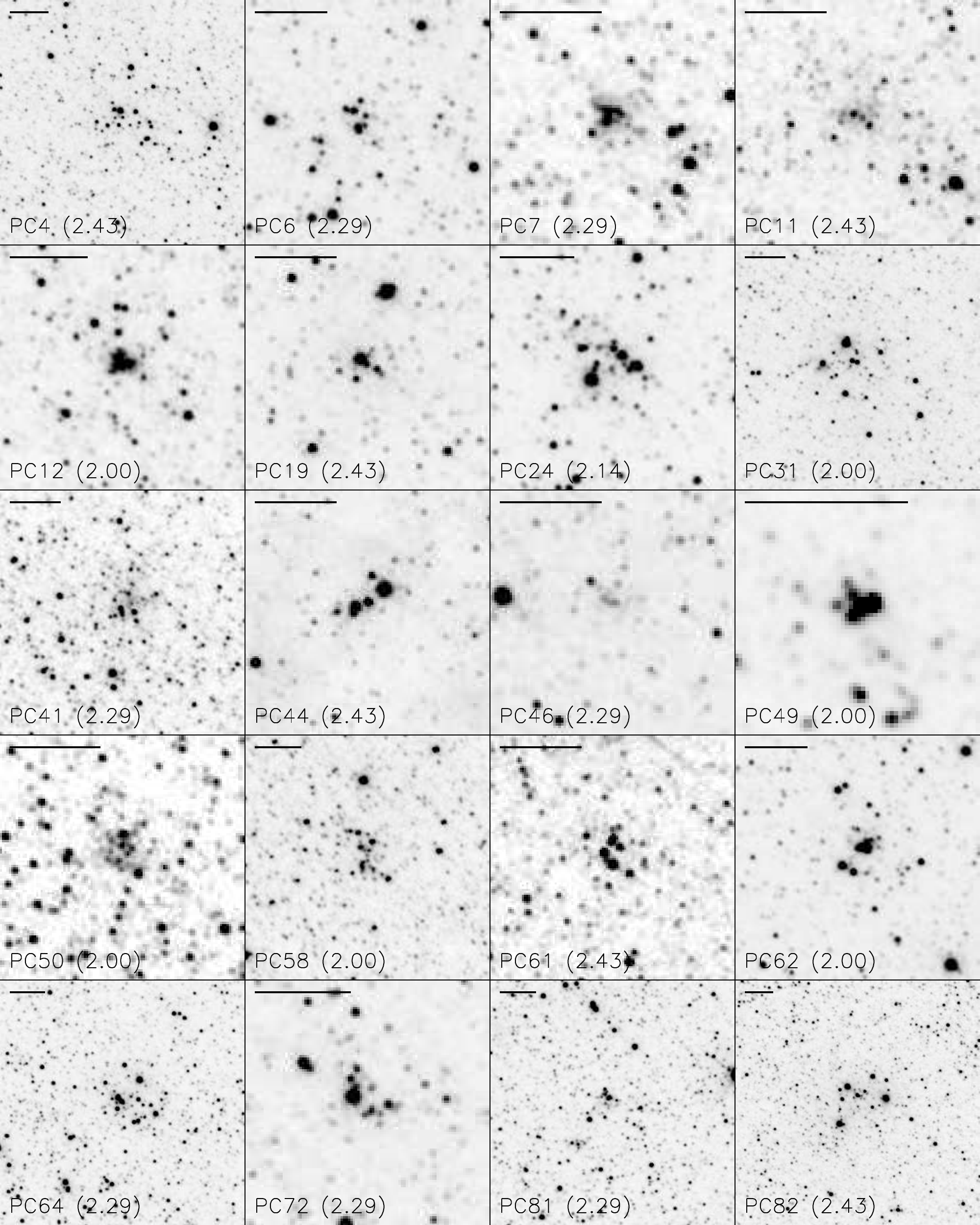}
\caption{Same as Figure \ref{cutout}, but for possible clusters listed in Table \ref{tbl2} with $2.0 \leq S_{by-eye} < 2.5$.  Figures \ref{cutout2}.1--\ref{cutout2}.12 are available in the online version of the Journal.}
\label{cutout2}
\end{figure}

The Year 1 cluster search yielded a catalog of 601 high-scoring clusters.  Table~\ref{tbl1} presents positions for each object, as well as other descriptive information (radii, photometric measurements) that will be described in Section~\ref{phot}.  Cutout images for each object are presented in Fig.~\ref{cutout}.  In addition, the spatial distribution of the clusters are shown in Fig.~\ref{footprint}.  Tabulated information and image cutouts for 237 possible clusters are presented in Table~\ref{tbl2} and Fig.~\ref{cutout2}, respectively.  Information about the possible clusters is provided for completeness, but due to the uncertain nature of their classifications, we exclude these objects from further analysis.

During the cluster search, we also identified 370 putative background galaxies.  We did not explicitly search for these objects, therefore this catalog does not constitute a complete sample of objects.  However, the potential usefulness of these identifications (e.g., as astrometric references, multi-wavelength source catalog cross-correlation) warrants its release.  We present this catalog of objects in Appendix \ref{bckgal}.

\subsection{Catalog Completeness} \label{comp}

To characterize the completeness of our cluster sample, we conduct artificial cluster tests that mimic the selection procedure of the clusters.  We use artificial clusters that span the range of ages and masses we expect to find in the cluster sample, while the size distribution is chosen to sample the minimum, average, and maximum sizes of the true sample (see Section~\ref{size}).  Ages and masses are chosen from a logarithmic grid of values, while sizes are drawn from the set of three characteristic values.  Specifically, we select ages ranging from 4 million to 10 billion years, masses ranging from $10^2$ to $10^5$ M$_\odot$, and profiles that have effective radii ($R_{eff}$; equivalently, half-light radii) of 1, 3 or 7~pc (0.26, 0.79, or 1.84~arcsec).

We create artificial clusters by populating a Padova isochrone \citep{Girardi10} of the appropriate age using a \citet{Kroupa01} stellar initial mass function.  Next, the stars are spatially distributed according to a \citet{King62} profile, assuming no mass segregation.  Finally, the size of the cluster and magnitudes of the individual stars are scaled appropriately to account for the distance of M31, assuming zero Galactic foreground or other internal M31 extinction.  For each test field, we insert 46 artificial clusters into individual raw (FLT) images using new functionality developed for the DOLPHOT photometry package.  The clusters are randomly positioned within each image to prevent search bias (as would result from a regular grid pattern), though we ensure that clusters are well-separated within the image so that they do not overlap or interfere with the photometry of other artificial clusters.  We drizzle the resulting images together to create searchable images in the optical F475W and F814W passbands.  In all, we created three fields of artificial clusters for each of the four full bricks in the Year 1 dataset, resulting in a total of $\sim$550 simulated objects.  In addition to the completeness tests that follow, these artificial clusters are also used for quality assurance of our photometric results, as discussed in Section~\ref{photsim}.

\begin{figure*}
\centering
\includegraphics[scale=0.55]{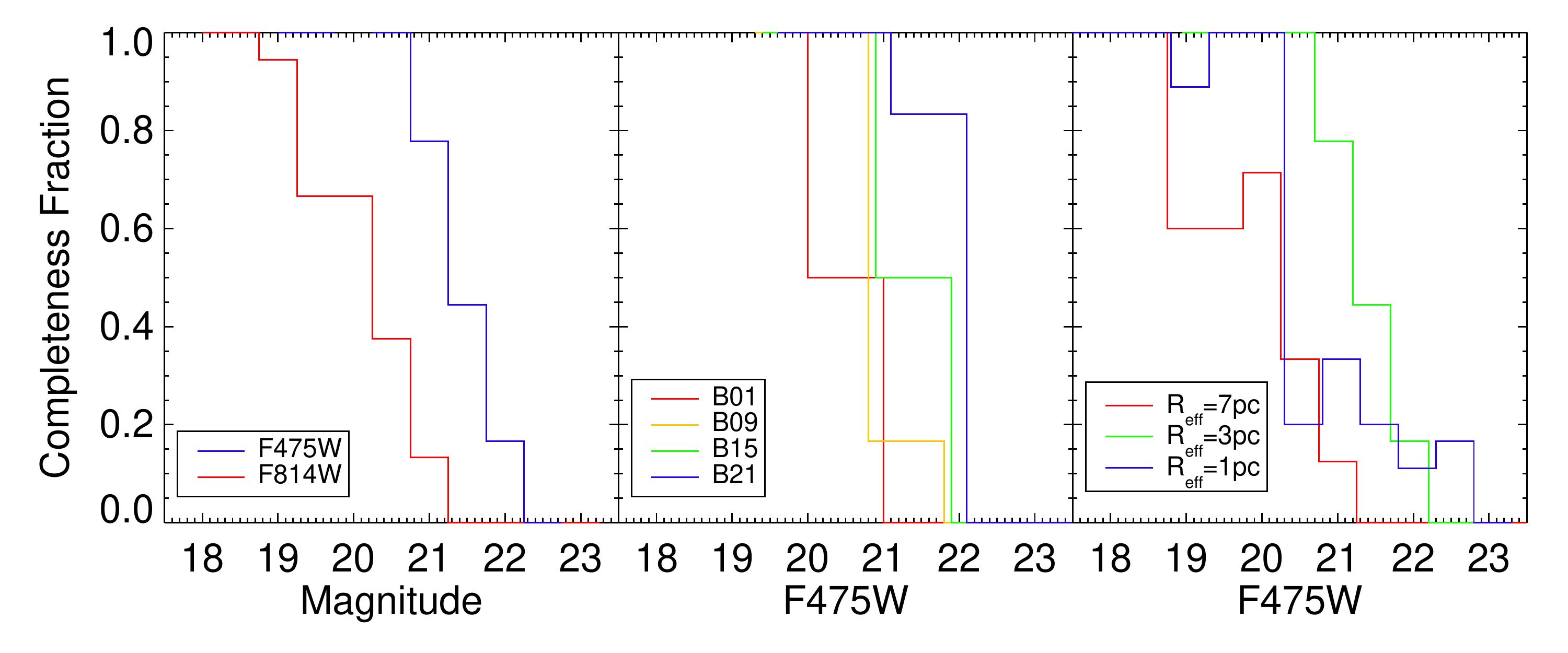}
\caption{Cluster completeness derived from artificial cluster tests.  Left: Completeness in the two optical filters, derived from 3 pc artificial clusters inserted in B09, B15, and B21.  Center: Variation in completeness as a function of brick membership, derived from $R_{eff}=3$ pc artificial clusters.  Right: Variation in completeness as a function of $R_{eff}$, derived from artificial clusters inserted in B09, B15, and B21.}
\label{completeness}
\end{figure*}

After creating the artificial cluster images, we identify clusters in the same way we searched the original images, with all eight team members searching each field and rating the reliability of the clusters.  By comparing the resulting cluster identifications to the full list of inserted clusters, we can estimate the completeness limits of the sample based on various cluster input parameters; the results of this analysis are shown in Fig.~\ref{completeness}.  The left panel shows the best estimate of the sample's characteristic cluster recovery fraction, computed using a subsample of 110 ``typical'' simulated clusters with $R_{eff}$ of 3~pc distributed within the outer three bricks (B09, B15, \& B21).  We estimate a 50\% completeness limit of $m_{F475W} \sim$21.2.  Considering distance and Galactic foreground dust reddening \citep[$E(B-V)=0.062$;][]{Schlegel98}, this translates to an absolute magnitude limit of $M_{F475W} \sim$ -3.5.

The true completeness of the sample as a whole, however, is a complicated function of cluster luminosity, size, and location within M31.  Using the full complement of simulated clusters, we explore the dependence of completeness on the latter two cluster properties.  The middle panel of Fig.~\ref{completeness} shows that variations in galactocentric position result in a $\sim$1 magnitude difference in completeness limits.  Detection limits are brighter for clusters located in the bulge-dominated inner galaxy due to luminous, crowded background fields.  Cluster size also plays a role, in which increasing size at constant luminosity results in reduced surface brightness and lower detection efficiency.  The right panel of Fig.~\ref{completeness} shows that there is a $\sim$1 magnitude difference in completeness when comparing clusters with $R_{eff}$ of 1 versus 7~pc.  To account for these multiple dependencies, we will undertake a larger and more rigorous set of completeness tests in future work to better characterize the subtleties of the completeness function.

\section{Integrated Photometry} \label{phot}

We use aperture photometry to measure the six band integrated fluxes of each cluster.  Aperture photometry consists of two main analysis tasks: defining an aperture (center and size) and determining the local background flux level.  We build upon photometry procedures used in M31 by \citet{Barmby01} and \citet{Krienke07}, with several refinements related to the assessment of local background levels and accounting for light that lies outside the photometric aperture.  A detailed description of photometric analysis procedures and results is provided in Section \ref{photprocedure}, followed by artificial cluster validation analysis in Section~\ref{photsim}.  Photometry zeropoints for the ACS and WFC3 cameras were obtained from the STScI webpage\footnote{\url{http://www.stsci.edu/hst/acs/analysis/zeropoints} and \url{http://www.stsci.edu/hst/wfc3/phot\_zp\_lbn}} and are listed in Table~\ref{zeropoints}.  All photometry is presented in the Vega magnitude system using the native HST passbands; we do not perform passband conversions.

\subsection{Aperture Photometry Procedure} \label{photprocedure}

The first step in aperture photometry is to define an aperture center.  Centers are estimated by centroiding on a F475W image that is smoothed using a 0.3$''$ (6 pixel) FWHM Gaussian kernel.  For certain clusters, particularly low luminosity objects whose light is dominated by a small number of bright sources, the flux-weighted positions determined by the automated procedure do not always accurately reflect the cluster center.  For this reason, we visually inspect central positions and manually adjust incorrect determinations.

\begin{figure*}[ht!]
\centering
\includegraphics[scale=0.65]{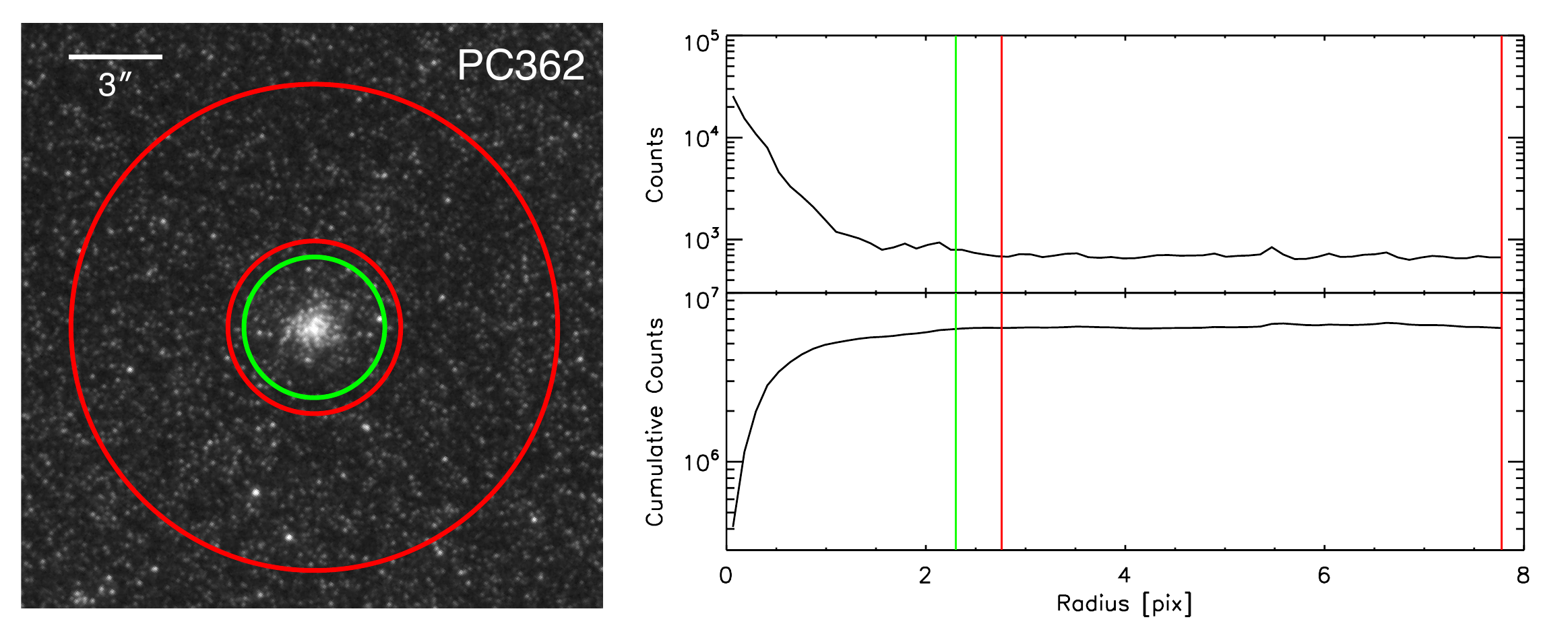}
\caption{Cluster aperture layout and growth curve for PC362.  Left: F475W image of cluster with aperture locations overplotted.  The green circle denotes the photometric aperture.  The red circles denote the inner and outer edges of the annulus used for background determination; this region is subdivided into ten equal-area annular subregions to estimate the variance in the background.  Right: The flux profile (top) and cumulative growth curve (bottom) for the cluster; vertical red and green lines correspond to the radii plotted in the left panel.}
\label{apexample}
\end{figure*}

We adopt a photometric aperture size that provides the largest signal-to-noise ratio for flux measurement by enclosing a maximum amount of cluster light while including as little background light as possible.  The clusters considered in this study vary by a factor of $\sim$10 in radius, and consequently, the chosen photometric aperture radii vary by the same amount.  We define circular apertures using growth curve analysis to determine an appropriate radius.  The aperture limit is defined at the radius where the cluster profile drops below the level of the noise in the background, equivalent to the point at which the curve of growth turns over and the increase in cumulative flux as a function of radius stops.  An illustrative example of a cluster image and growth curve is provided in Fig.~\ref{apexample}.  The aperture radii ($R_{ap}$) are determined by visual inspection of the growth curves for each cluster and are reported with the photometric results.  While it would be preferable to adopt an algorithmic approach for defining $R_{ap}$, the relatively noisy character of the local background significantly complicates automated determinations.  We perform aperture definition and growth curve analysis on the F475W image, which provides the best combination of signal-to-noise and contrast between cluster and field populations for a wide range of cluster ages.  Apertures of the same angular size are used for the five other images.  Photometric aperture radii range between 0.5$''$ and 6$''$ for the Year 1 sample.  

Aperture photometry depends greatly on the determination of the background flux level.  Following traditional photometric terminology, we also refer to non-cluster background light as the ``sky'' or the ``sky background''.  For the PHAT cluster sample, the sky is made up of two components: individual resolved stars and unresolved light.  Traditionally, the background is determined using the mode of sky region pixel values.  However, the fact that resolved stars are a true component of the background light calls for an alternate statistical treatment.

We define ten annular sky regions that encircle the photometric aperture, extending radially from $1.2 \times R_{ap}$ to $3.4 \times R_{ap}$.  Each of the annuli are equal in area to the photometric aperture, to accurately measure the dispersion of the sky background based on an equal number of pixel samples.  Figure~\ref{apexample} shows an example of the aperture layout, where we denote the inner and outer extent of the sky measurement annuli.  Next, we calculate the total integrated flux within each of the sky regions.  We perform iterative $2\sigma$ rejection on these ten fluxes, thereby excluding regions that contain bright stars or other objects not representative of the typical sky background.  We adopt the mean of the non-rejected sky fluxes as the sky background value, and the standard deviation of these fluxes as a measure of the uncertainty in the sky background determination.  We propagate the uncertainty from the sky level determination into the overall cluster photometry by adding it in quadrature with the cluster's flux measurement uncertainty.

In agreement with the HKC cluster studies, the uncertainty in the sky background determination dominates the overall uncertainty in cluster magnitudes.  When compared to previous ground-based cluster photometry in M31 or other HST-based extragalactic cluster photometry, our magnitude uncertainties appear larger.  Our inclusion of sky level uncertainty into the reported values account for these larger overall magnitude uncertainties.  Uncertainty in the determination of cluster fluxes, comparable to the errors reported in other cluster catalogs, never rise above 0.01 mag for any object in the Year 1 catalog.

\begin{figure*}[ht!]
\centering
\includegraphics[scale=0.65]{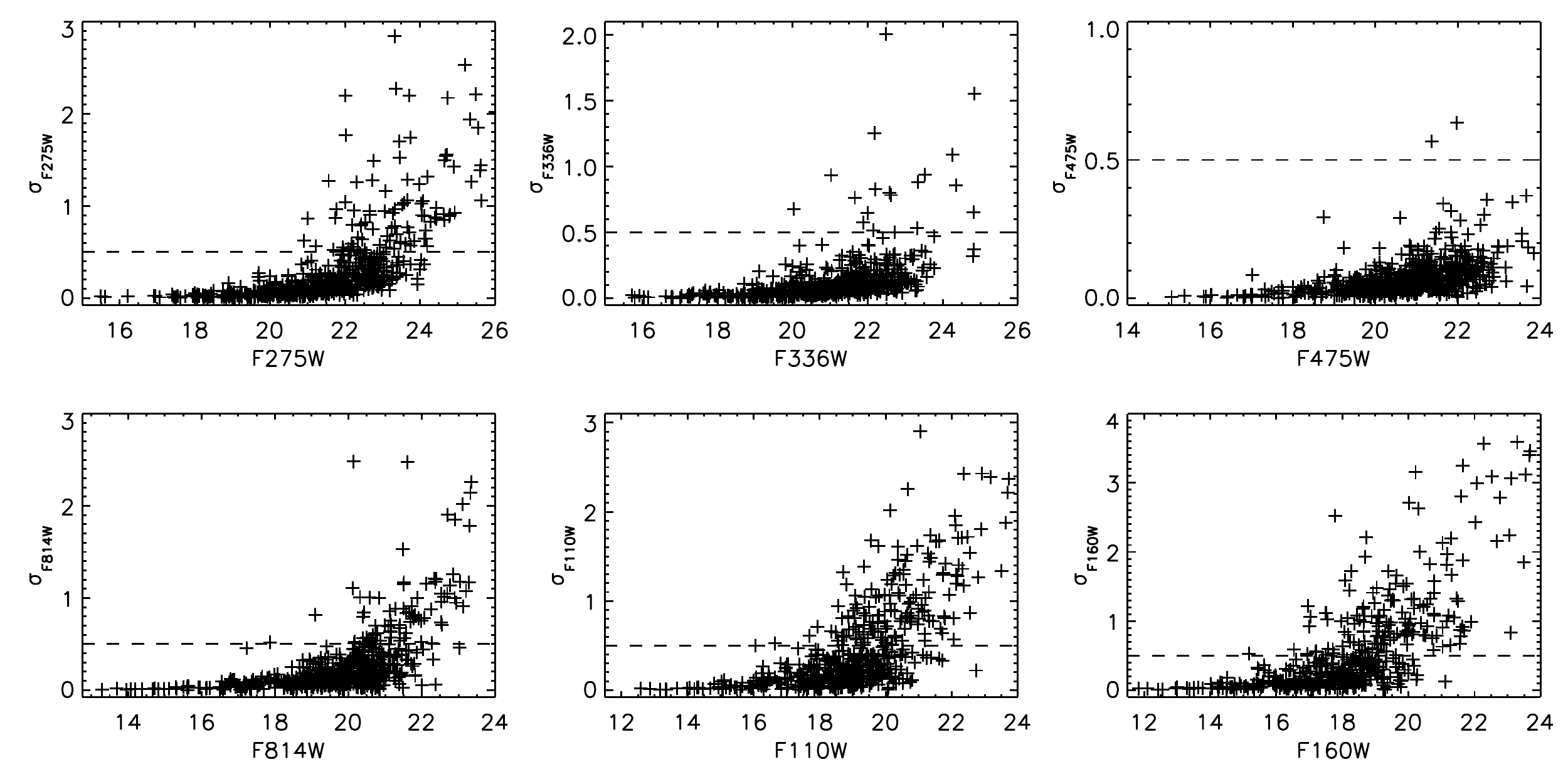}
\caption{Photometric errors for integrated cluster photometry in each of the six PHAT passbands.  The scale of the y-axis varies between panels.  A standard reference at $\sigma = 0.5$ mag is provided for comparison.}
\label{errors}
\end{figure*}

The resulting magnitudes measured within $R_{ap}$, hereafter referred to as \textit{aperture magnitudes}, are presented in Table~\ref{tbl1} for the cluster sample and Table~\ref{tbl2} for the possible cluster sample.  These aperture magnitudes represent high signal-to-noise, spatially matched measurements of cluster light, and are optimal for calculating cluster colors.  We plot the photometric uncertainties as a function of magnitude for the cluster sample in Fig.~\ref{errors}.  To assess relative reliability in the six passbands, we count the number of well-determined ($\sigma < 0.5$ mag) photometric measurements in each band, and record the results in Table~\ref{photdat}.  The F475W imaging provides the highest-quality measurements, as shown by that filter's small photometric uncertainties, followed by the F336W image which probes a similar wavelength regime.  The F814W image has increased levels of photometric error due to the reduced contrast between cluster and field populations.  The quality of the measurements is lower in the three remaining filters due to intrinsic wavelength-dependent limitations; the F275W measurements suffer from low signal-to-noise for all but the youngest, bluest clusters, while the F110W and F160W measurements suffer from high sky background levels and poor cluster-field contrast.  In addition to the problem of faint signal, we note that $\sim$10\% of F275W magnitudes (and F336W magnitudes, but at a lower level) are affected by cosmic ray artifacts.  While a vast majority of these defects are adequately corrected for by our image processing, we caution that a few percent of the UV magnitudes might still be adversely affected.

In addition to the aperture magnitudes, we also provide estimates of the effective radius ($R_{eff}$; equivalently, the half-light radius) for each cluster in Tables \ref{tbl1} and \ref{tbl2}.  These estimates are obtained by measuring the cluster flux profile and interpolating this curve to find the radius that contains half of the light within $R_{ap}$.  We use the F475W cluster light profiles to make these assessments, given their good data quality.  We discuss the resulting sizes in Section~\ref{size} and use them to calculate aperture corrections in the next section.  We recognize that the $R_{eff}$ estimates will systematically underestimate the true cluster sizes due to the fact that they are calculated using aperture magnitudes that fall short of measuring the full luminosity of the cluster.  However, we expect the impact of this underestimation to be small due to the steepness of the cluster light profile.

\subsubsection{Aperture Corrections} \label{apcor}

The aperture magnitudes presented above measure a majority of the cluster light.  However, these values do not account for light that lies below the noise level of the sky background, beyond the limits of our photometric aperture in the faint outer wings of the cluster profile.  We correct for this missing flux by using $R_{eff}$ determinations to approximate the cluster's luminosity profile shape, then extrapolate this profile past the limits of our photometric aperture to make an estimate of the cluster's total light.  The magnitude difference calculated between the original aperture magnitudes and this total light estimate are equivalent to an aperture correction.

To make this profile-based extrapolation, we require a cluster profile shape and a normalization for that profile.  We adopt a \citet{King62} profile, assuming a concentration ($c=R_{tidal}/R_{core}$) that matches the characteristic profile of PHAT clusters, as obtained during preliminary cluster profile fitting ($c=7$).  Next, we use a cluster's $R_{eff}$ to set the radial scaling of that characteristic King profile.  Finally, we normalize the scaled profile such that the integrated flux within $R_{ap}$ matches the cluster's aperture magnitude measurement.  Once normalized, we calculate the fraction of flux that lies outside $R_{ap}$ and transform this value into an aperture correction in magnitudes, which is presented in Tables \ref{tbl1} and \ref{tbl2} for all clusters.  These corrections may be applied to the aperture magnitudes to obtain \textit{total magnitudes}\footnote{Total Magnitude = Aperture Magnitude + Aperture Correction}.  These total magnitudes are optimal for the estimation of total cluster magnitudes and luminosities.  While the aperture corrections were derived in the F475W passband, they may be used for all filters under the simplifying assumption of flat radial color profiles in the outer parts of the cluster.

\begin{figure}[ht!]
\centering
\includegraphics[scale=0.5]{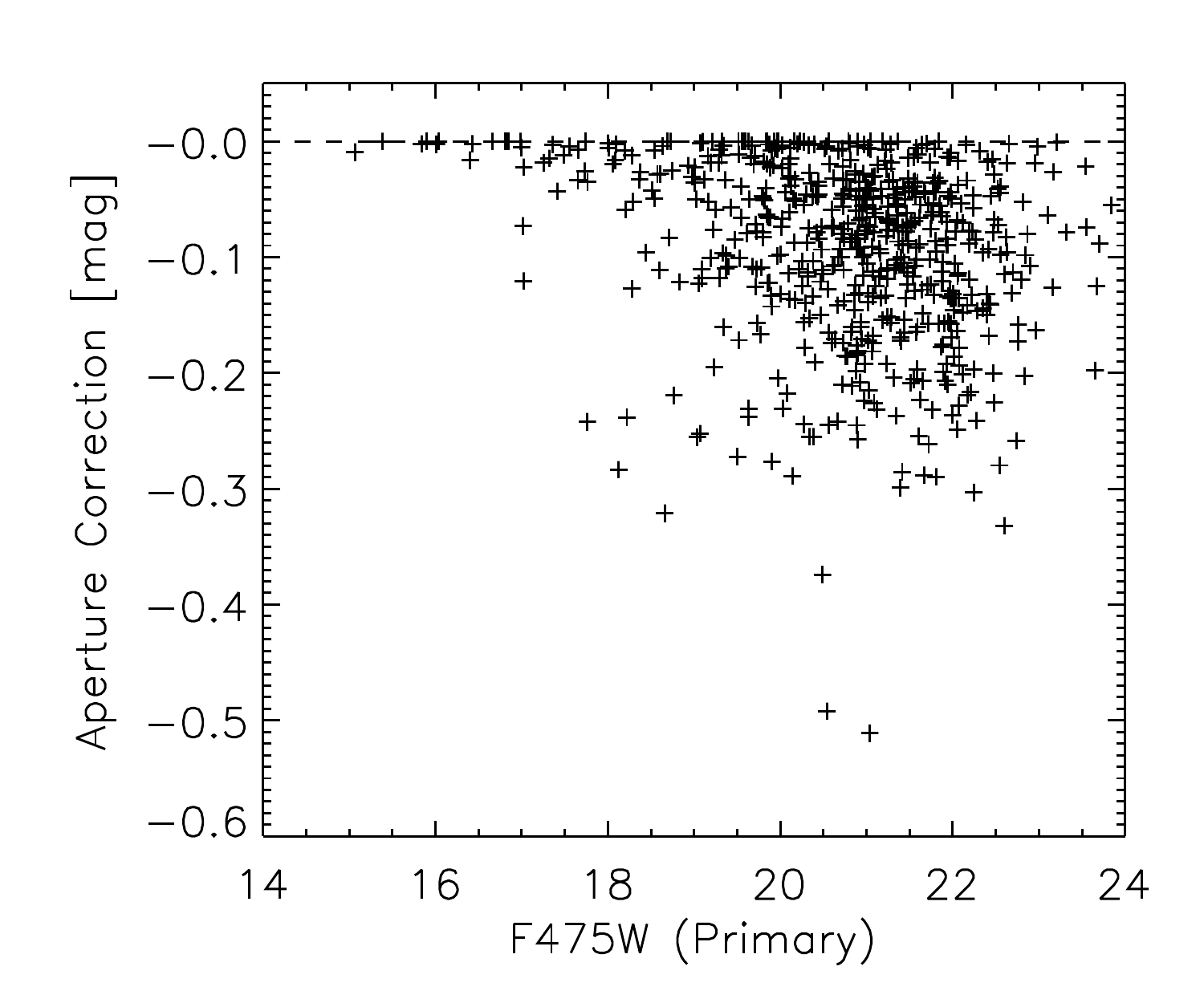}
\caption{Aperture corrections, used for converting aperture magnitudes to total magnitudes.  The corrections are derived from cluster profile extrapolation.}
\label{apcor_data}
\end{figure}

The amplitude of the aperture corrections are presented in Fig.~\ref{apcor_data}.  Over the sample of clusters, the corrections vary from 0.0 to -0.6 mag, with a median correction of -0.1 mag.  These corrections are negligible for the brightest clusters, where a majority of the light is detectable above the noise level of the sky background.  The corrections become larger for fainter clusters, due to their low cluster-to-field flux contrast.  The simplifying assumption of a universal cluster profile shape that varies only as a function of $R_{eff}$ provides suitable accuracy for this correction, as shown by artificial cluster tests that follow in Section~\ref{photsim}.  Full cluster profile and structural parameter fitting is currently underway (Fouesneau et al. 2012, in prep.), and these results could be used to further refine these profile extrapolations and improve upon our estimates of total cluster light in future work.  However, we expect little overall improvement in photometry as a result of increased aperture correction precision because these corrections are comparable in size to the amplitude of the photometric uncertainties for most clusters.

\subsection{Artificial Cluster Photometry Experiments} \label{photsim}

We use artificial clusters to assess the uncertainties and biases associated with our photometry procedures.  For these tests, we only consider clusters that lie in the outer three bricks, have $R_{eff}$ of 3~pc, and were detected as part of our artificial cluster search in Section~\ref{comp}.  Analysis of this sample of 76 artificial clusters provides an evaluation of photometric accuracy for ``typical'' clusters in the Year 1 sample.  We process the simulated objects using photometric procedures identical to those described in the previous section.

\begin{figure}
\centering
\includegraphics[scale=0.58]{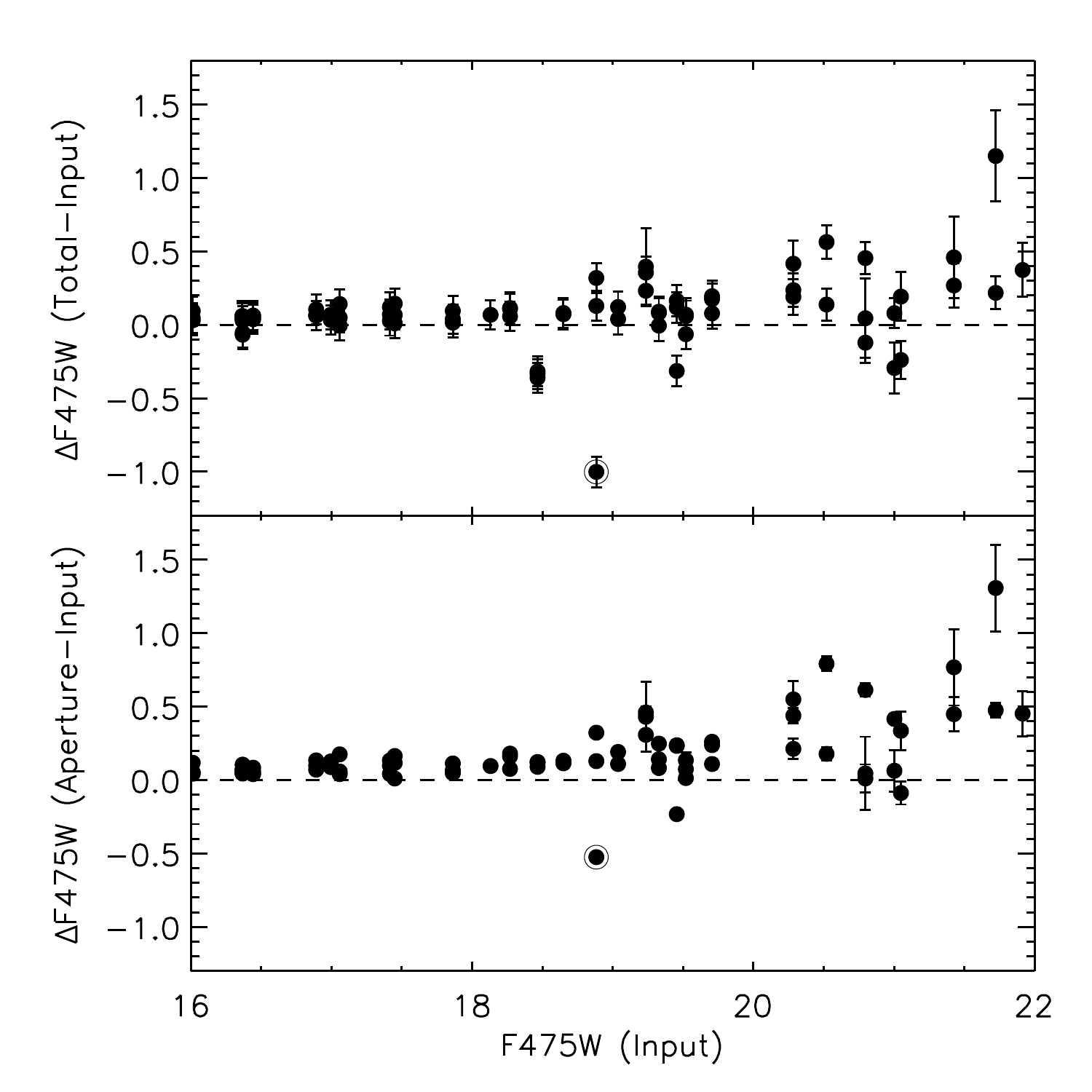}
\caption{F475W magnitude differences between input and recovered magnitudes for a subset of simulated clusters inserted into B09, B15, and B21, adopting a $R_{eff}$ of 3 pc.  Top: Differences between the measured total magnitudes and the input magnitudes.  Bottom: Differences between the measured aperture magnitudes and the input magnitudes.  The lowest outlier in both panels (circled) differs due to a bright (F475W$\sim$19) nearby field star that overlaps the cluster.}
\label{faketst}
\end{figure}

The results of these tests are shown in Fig.~\ref{faketst}.  The artificial cluster tests show good agreement ($\Delta$F475W $\lesssim$0.1 mag) between the input and output magnitudes for bright clusters (F475W $< 19$).  For less luminous clusters, aperture magnitudes are fainter than input magnitudes by up to $\sim$0.4 mag at F475W $\sim 21.5$.  However, total magnitudes are more successful in recovering input magnitudes accurately, showing a smaller faintward bias of $\sim$0.2 mag at F475W $\sim 21.5$.  This bias represents a $\sim$0.5$\sigma$ deviation when compared to the $\sim$0.4 mag scatter in the photometry.  Finally, we acknowledge that these photometric experiments are slightly idealized, for instance due to our consideration of only a single family of cluster profile shapes.  However, this testing confirms that our photometry techniques provide accurate assessments of cluster properties and their associated uncertainties.

\section{Comparison to Existing M31 Cluster Studies} \label{oldcat}

As discussed in the introduction, there is a long history of stellar cluster studies in the Andromeda galaxy.  Decades of effort have produced a wealth of knowledge on this topic.  To place our findings in context, in this section we cross-reference our cluster identifications with existing catalogs, allowing us to reference previous work on the same objects and compare the results of our cluster analysis to existing catalogs.

To compile a list of known clusters located within the Year 1 footprint, we began by searching the Revised Bologna Catalog\footnote{\url{http://www.bo.astro.it/M31/}} \citep[RBC;][last updated 2009 December to v4.0]{Galleti04}.  This excellent resource has aggregated all known cluster identifications from early catalogs \citep[e.g.,][among many others]{Vetesnik62, Sargent77, Crampton85, Battistini87, Battistini93, Barmby00} as well as more recent works \citep[e.g.,][]{Kim07, Huxor08, Caldwell09}.  Other than the RBC, we searched the HKC and other works published since the most recent RBC revision \citep{Vansevicius09, Peacock10, Fan10}.  Our search of the HKC produced additional objects for cross-matching, however no new objects were recovered from the three other catalogs.  The \citet{Peacock10} catalog is composed solely of edits and reclassifications from an earlier version of the RBC (v3.5).  Similarly, the objects studied in \citet{Fan10} are derived directly from v4.0 of the RBC.  Finally, we find no overlap between any part of the PHAT footprint (existing or planned coverage) and the southwest region of M31 studied by \citet{Vansevicius09}.  In the discussion that follows, we consider cluster classifications with respect to those provided in the RBC v4.0.

\begin{figure}
\centering
\includegraphics[scale=0.45]{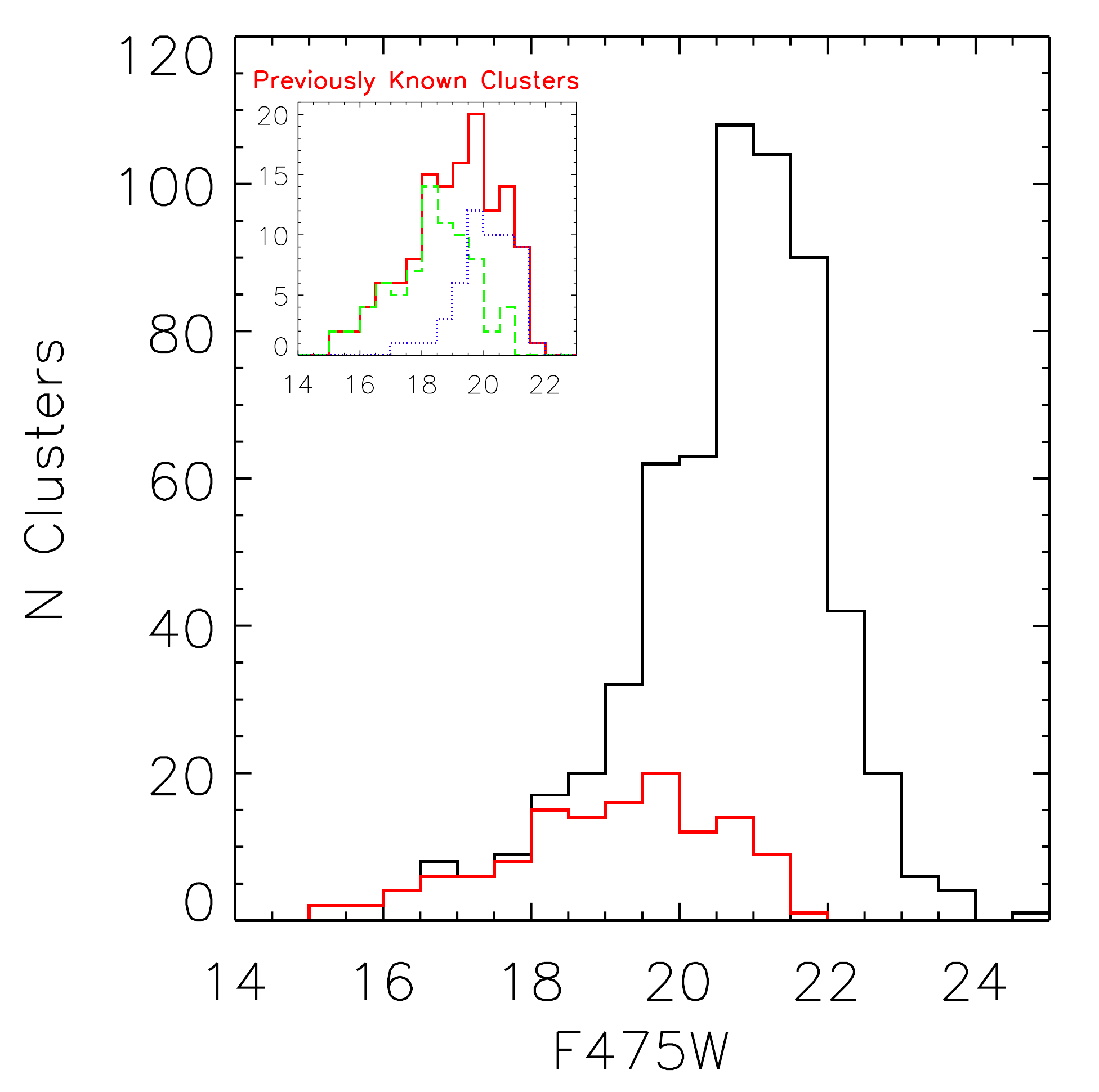}
\caption{Magnitude histogram comparing the Year 1 PHAT clusters (black) to objects from existing catalogs that fall within the Year 1 survey footprint (red).  The inset shows a breakdown of the previously known clusters into objects that were identified using ground-based imaging (green) and those that were identified using HST imaging (blue).}
\label{histogram}
\end{figure}

The RBC and HKC contain a total of 146 published clusters, 32 cluster candidates, and 84 other non-cluster classifications (foreground stars, background galaxies, and \hii\ regions) that lie within the Year 1 PHAT survey footprint.  We cross-match these previously known objects with all identifications made as part of the Year 1 search (cluster, possible clusters, and unlikely objects).  Further, we re-examined the PHAT data at positions of previously cataloged objects that were not matched to PHAT candidates to ensure the complete re-analysis of all existing catalog entries that lie within the Year 1 footprint.  In total, we classify 132 of the previously known objects as clusters, 12 as possible clusters, and reject the 118 remaining objects.  We note that all confirmed Year 1 PHAT clusters were identified independently as part of our blind cluster search.  Tables \ref{tbl1} and \ref{tbl2} provide cross-matched names of clusters using the naming convention of \citet{Barmby00}, consisting of the identifier from the Revised Bologna Catalog, followed by the identifier of the next most significant cluster catalog, where such exists.  Additional details concerning the comparison and reclassification of existing cluster identifications is provided in Appendix \ref{revise}.  There we provide catalog-specific commentary on the validity of the previously published classifications, as well as object-by-object classification revisions for the RBC and the HKC.

A comparison between ground-based and space-based M31 cluster catalogs reveals the importance of high spatial resolution imaging in cluster identification work.  In Fig.~\ref{histogram}, we present a histogram of the number of cluster as a function of apparent magnitude for the Year 1 cluster sample of 601 clusters and the 132 previously known objects confirmed as clusters in the PHAT data.  It is immediately apparent from the plot that HST imaging has allowed the PHAT cluster survey to identify hundreds of low luminosity clusters that could not be identified from ground-based data.  As shown in the inset of Fig.~\ref{histogram}, the sample completeness associated with previously known clusters discovered from ground-based data (green dashed histogram) drops precipitously at $m_{F475W} > 18$ ($M_{F475W} > -6.5$).  This fall-off in completeness reflects the difficulty in differentiating between single unresolved stars and compact clusters in low resolution images.  In contrast, high spatial resolution imaging from HST enables the identification of clusters 2-3 magnitudes fainter than previous ground-based surveys.  Within this fainter luminosity range, Fig.~\ref{histogram} shows an order of magnitude increase in the number of objects identified in the PHAT cluster catalog when compared to previous HST-based cluster survey work (HKC; blue dotted histogram in inset).  This improvement results from the order of magnitude increase in spatial coverage provided by PHAT when compared to the limited number of previous, targeted HST observations that fall within the Year 1 survey footprint (35 $\text{arcmin}^2$ versus 390 $\text{arcmin}^2$ in PHAT Year 1).

In addition to the catalog comparison presented here, in Appendix~\ref{photcompare} we compare photometry results presented in this work to those of existing catalogs.  This analysis acts as quality assurance for the photometry presented here, and provides the reader with an assessment of the inherent differences between the sets of photometric results.

\section{Year 1 Clusters: Photometric Properties} \label{results}

The Year 1 cluster sample, derived from $\sim$1/4 of the total expected PHAT survey data, provides the first glimpse of what can be expected from the full balance of the PHAT stellar cluster survey.  In Section \ref{oldcat}, we showed that our catalog represents a considerable increase in the number and diversity of clusters known in M31.  Our excellent HST-based imaging should lead to factor of $>$4 increases in the number of known clusters within the PHAT survey footprint.

\begin{figure*}
\centering
\includegraphics[scale=0.35]{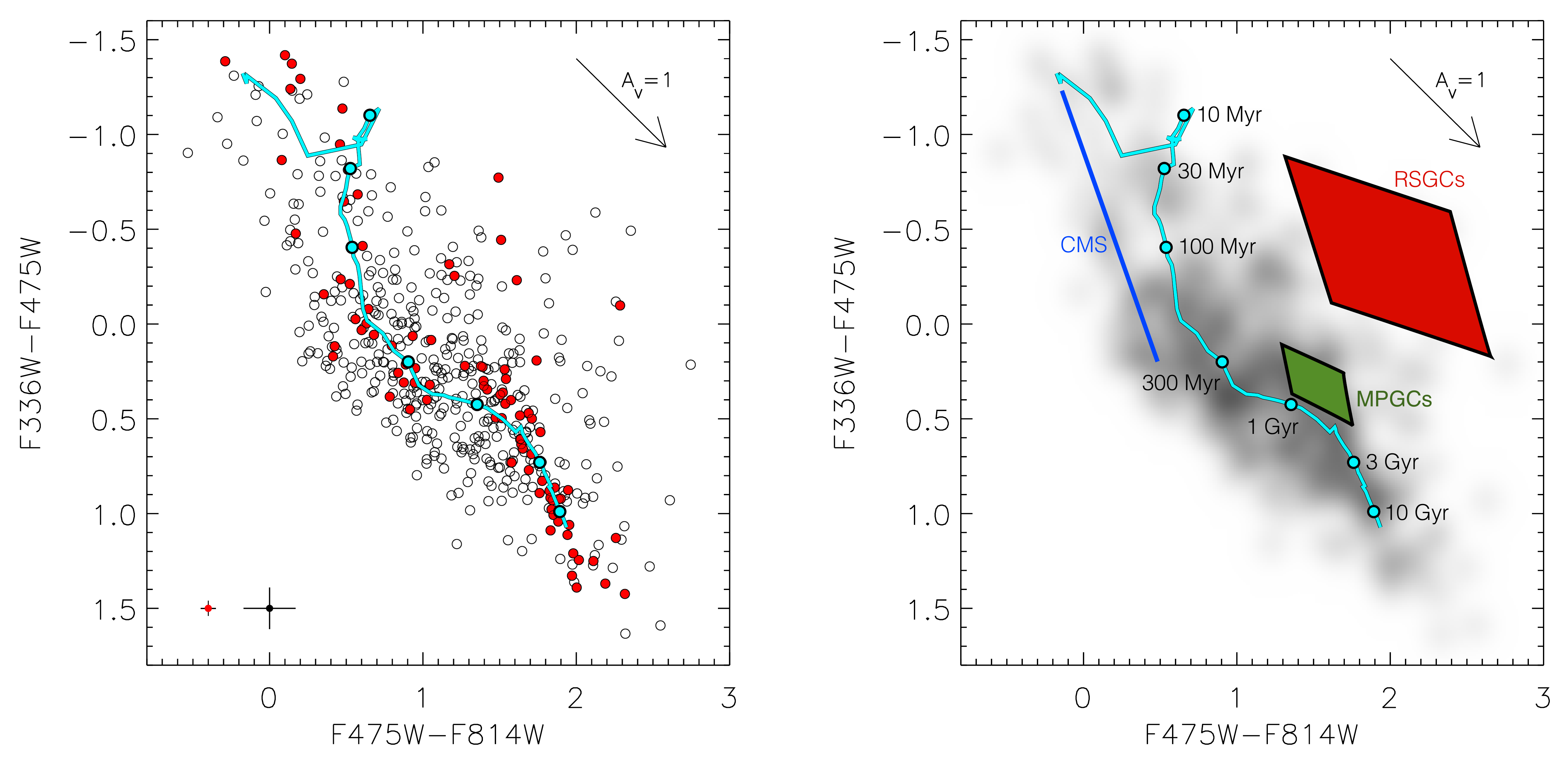}
\caption{Left: Color-color diagram for the 482 Year 1 clusters with well-constrained ($\sigma< 0.5$ mag) photometry in the F336W, F475W, and F814W passbands.  A subset of the 92 most luminous clusters (F475W $< 19.5$) are highlighted in red.  Padova SSP models for Solar metallicity, reddened to account for Galactic foreground extinction, are plotted as a cyan line for reference.  Cyan points act as age indicators, spaced at 0.5 dex increments beginning at 10 Myr.  Median error bars for the bright (red) and faint (black) cluster samples are displayed in the lower left corner.  Right: A smoothed, grayscale version of the color-color diagram, with SSP models shown again as reference.  The red box denotes the location of anomalous red supergiant clusters (RSGCs), which contain luminous evolved stars that strongly affect the cluster's integrated colors.  The green box denotes the parameter space populated by metal poor globular clusters (MPGCs).  The blue line denotes the modeled location of the cluster main sequence (CMS), reddened by $A_V$=0.4 mag to match the color distribution of the Year 1 cluster sample.  The CMS represents the color-color sequence populated by low mass clusters that host no evolved stars due to stochastic sampling of the cluster's stellar mass function.}
\label{colorcolor}
\end{figure*}

To obtain a better sense for the type of clusters we have identified in the Year 1 sample, we present color and magnitude distributions from the catalog photometry.  For this analysis, we select a subset of objects with well-determined photometry, where the uncertainties in the F336W, F475W, and F814W magnitudes are each less than 0.5 mag.  This quality cut results in a subsample of 482 well-characterized objects that we use to explore the properties of the catalog.

We plot a color-color diagram for clusters with well-determined photometry in Fig.~\ref{colorcolor}.  This diagram aids our ability to assess the cluster age distribution.  To provide reference points to guide the eye, we overlay the stellar evolution model predictions from the Padova group \citep{Girardi10}.  These model tracks assume solar metallicity and are reddened to account for foreground Galactic extinction \citep[$E(B-V)=0.062$;][]{Schlegel98}.  On this plot, ages increase as we follow the evolutionary track from the upper left to the bottom right.  In addition to the initial cluster selection based on photometric uncertainties, we define a second subset of clusters based on cluster luminosity.  The red points denote the well-determined subsample's 92 most luminous clusters, with F475W $< 19.5$.

The first conclusion we draw from Fig.~\ref{colorcolor} is that the Year 1 cluster sample includes a wide range of ages.  Clusters populate the full length of the model evolutionary track, with a large number of objects populating an intermediate age range (300 Myr to 3 Gyr).  However, the mapping from position in the color-color diagram to age suffers from well-known degeneracies with extinction and metallicity.  For example, old (12-14 Gyr) metal-poor ([Fe/H] $\lesssim -1.0$) globular clusters inhabit the same position on this diagram as $\sim$100 Myr old, solar metallicity clusters with $A_V$ of $\sim$1.5 mag.  The green region in the right panel of the figure denotes the shared color-color region where this particular age-metallicity degeneracy exists.

Second, we observe that the cluster color distribution is affected by the effects of stochastic stellar mass function sampling in low-mass ($<10^4$ \solmass) clusters.  The red region in the right panel of Fig.~\ref{colorcolor} highlights color outliers that have anomalous red F475W-F814W colors.  While young clusters ($<$50 Myr) suffering from large amounts of dust attenuation ($A_V >$2 mag) can populate this region of the diagram, these red colors are more frequently caused by the presence of a small number of bright evolved supergiant stars that can bias integrated cluster colors.  We discuss this behavior in greater detail in Section~\ref{rsg}.  In addition to the red outliers, stochastic effects can also cause integrated cluster colors to appear bluer than model predictions.  The fluctuation in the small number of evolved stars sometimes results in the complete absence of supergiant cluster members, meaning that the cluster's integrated light is emitted exclusively by main sequence stars.  Such clusters fall onto a linear sequence in color-color space we refer to as the cluster main sequence.  We highlight this feature in blue in the right panel of Fig.~\ref{colorcolor}.

\begin{figure}
\centering
\includegraphics[scale=0.45]{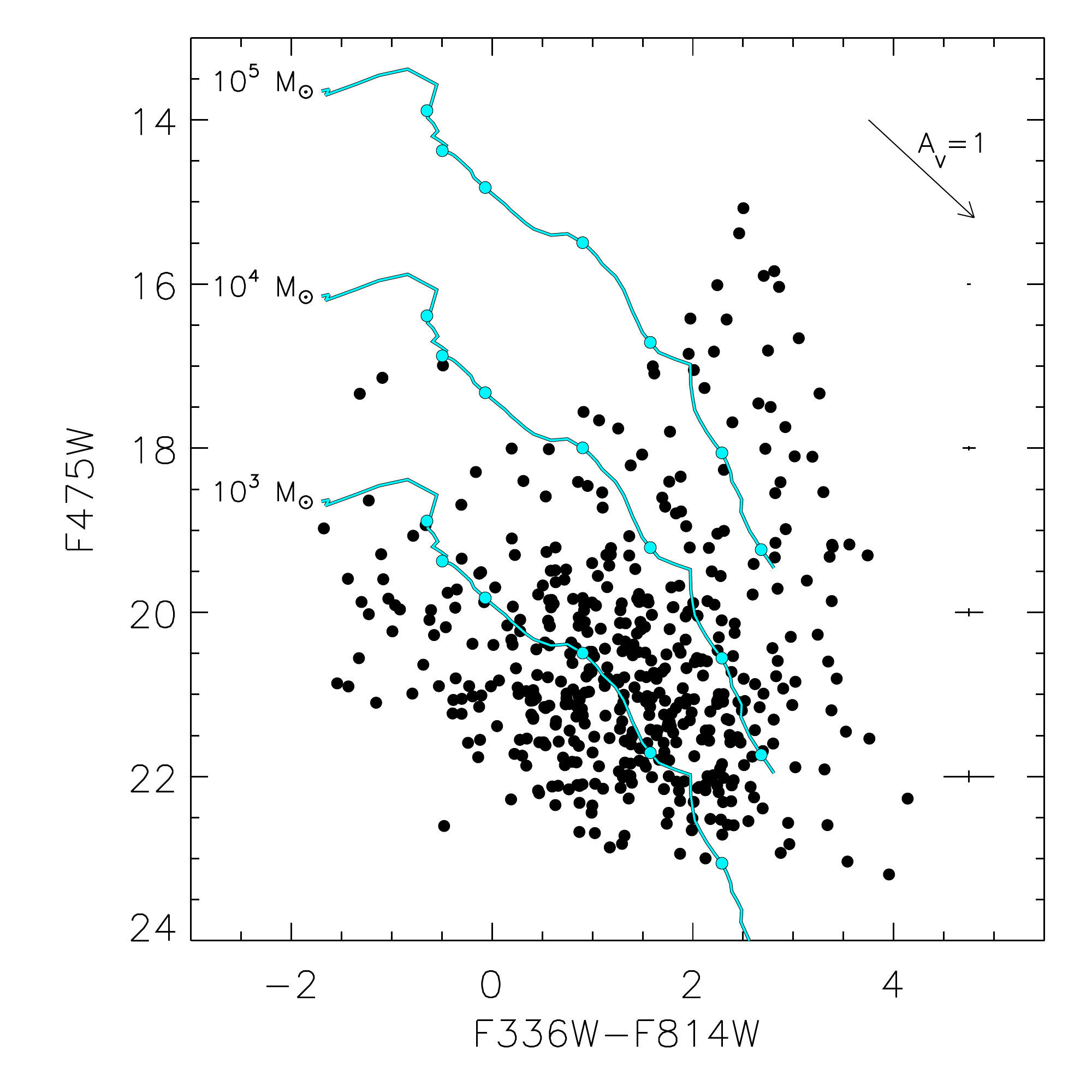}
\caption{Color-magnitude plot for the 482 Year 1 clusters with well-constrained ($\sigma< 0.5$ mag) photometry in F336W, F475W, and F814W passbands.  Padova SSP models for three cluster masses ($10^{5}$, $10^{4}$, $10^{3}$ M$_\odot$) at Solar metallicity are plotted for reference, with age indicators spaced at 0.5 dex increments beginning at 10 Myr.  Characteristic median error bars are shown on the right side of the plot, where each point describes the uncertainties of points within a 2 mag bin in luminosity.}
\label{colormag}
\end{figure}

Next, we plot a cluster color-magnitude diagram in Fig.~\ref{colormag} to assess the cluster mass range probed by the Year 1 sample.  As in Fig.~\ref{colorcolor}, we plot foreground reddened, solar metallicity Padova stellar models for reference.  The PHAT clusters span $\sim$8 mag in F475W luminosity, translating to $>$3 orders of magnitude in cluster mass.  This range indicates that the PHAT cluster sample hosts a wide variety of clusters, spanning systems that contain hundreds of solar masses up to those with a million solar masses.  The most luminous clusters, however, all appear to have red colors (F336W-F814W $\sim$ 2.5), forming a vertical sequence of objects on the right side of the plot.  This results from the fact that most massive ($> 10^{5}$ \solmass) clusters in the Year 1 sample are old globular clusters.

Figures \ref{colorcolor} and \ref{colormag} show that the sample of clusters assembled from the PHAT dataset provide a top-to-bottom assessment of the M31 cluster population.  Few datasets have the ability to sample objects across a variety of stages in cluster evolution over such a large, uninterrupted mass range.  This diversity makes the PHAT cluster sample a valuable tool to better understand the formation and dissolution of star clusters.

\section{Discussion} \label{discuss}

\subsection{Luminosity Functions} \label{lum}

\begin{figure*}[ht!]
\centering
\includegraphics[scale=0.5]{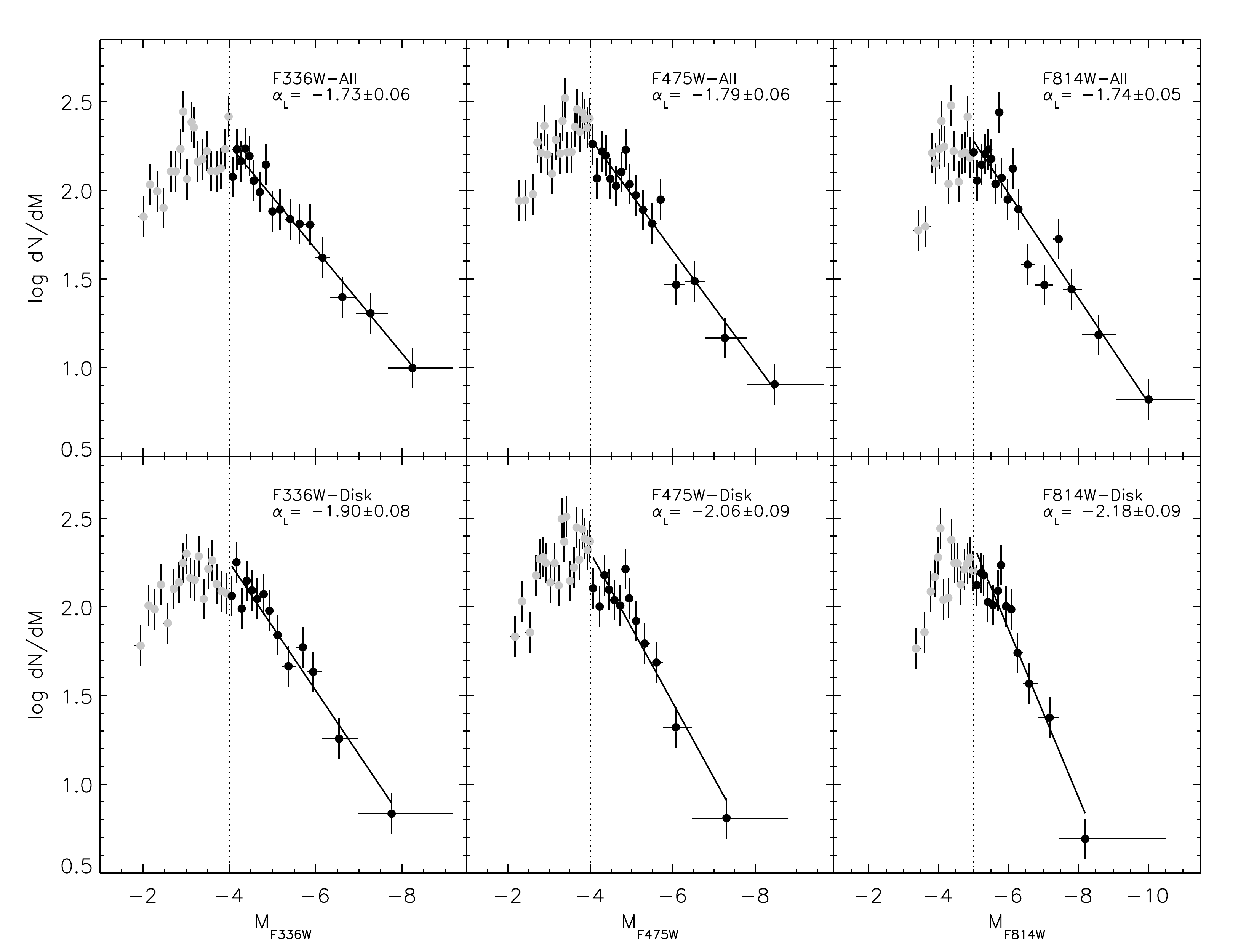}
\caption{Luminosity functions for the F336W (left), F475W (center), and F814W (right) passbands.  Top panels display the luminosity functions for clusters drawn from the complete Year 1 sample, and the bottom plots show the luminosity functions considering only disk clusters that lie outside the bulge (i.e., excluding B01 clusters).  Only clusters with well-constrained ($\sigma< 0.5$ mag) photometry are included in the analysis.  Clusters have been corrected for Galactic foreground dust attenuation, but not attenuation within M31.  Data is plotted using variable binning, such that each point represent an equal number of clusters (N=15).  Vertical error bars represent Poisson uncertainties due to the number of objects per bin, and horizontal error bars represent the magnitude range of each bin.  The luminosity function slope is fit down to completeness limits denoted by vertical dotted lines at -4.0 for F336W and F475W, and -5.0 for F814W.}
\label{lumfun}
\end{figure*}

\begin{figure*}[ht!]
\centering
\includegraphics[scale=0.5]{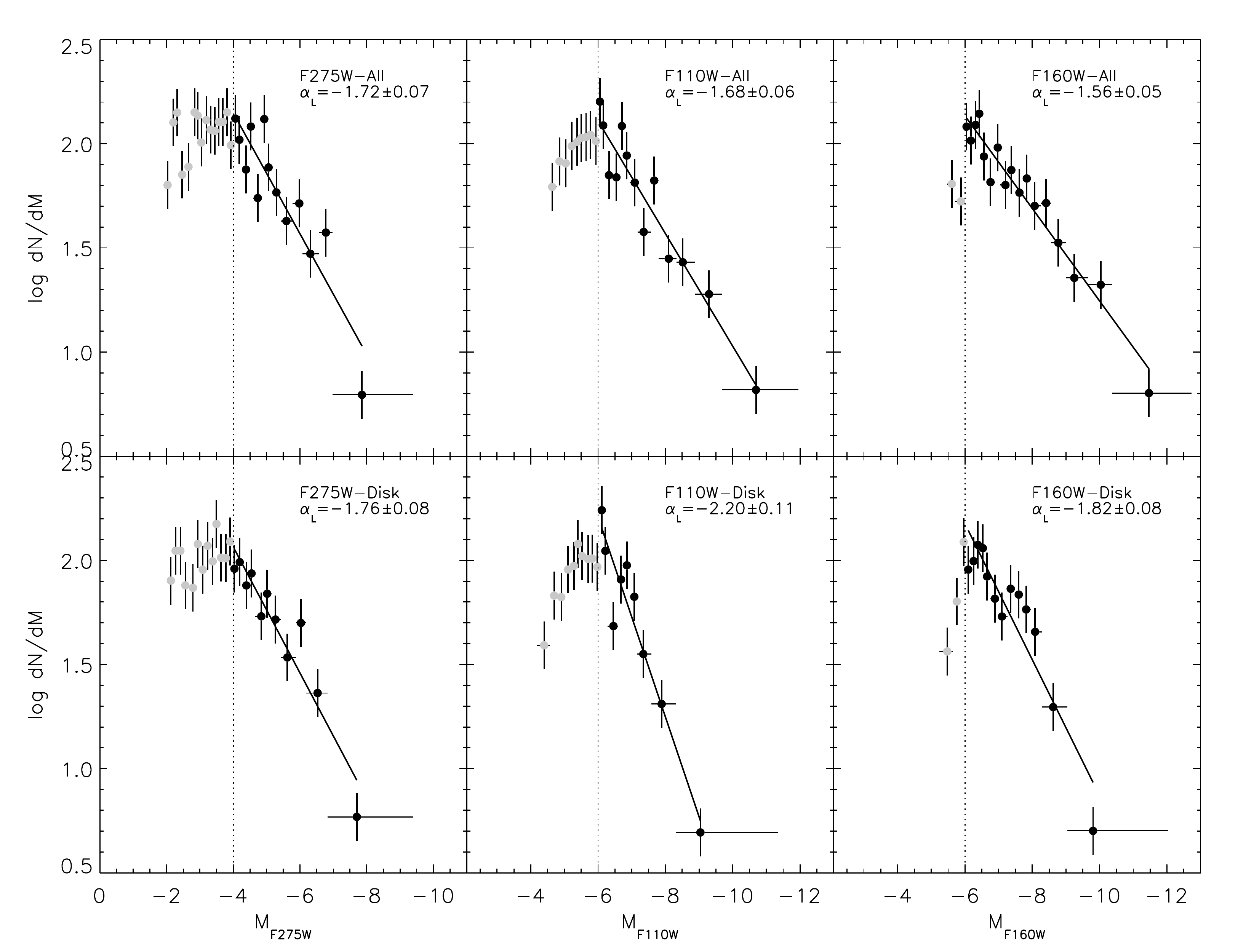}
\caption{Same as Figure \ref{lumfun}, but for the F275W (left), F110W (center), and F160W (right) passbands.  Completeness limits are -4.0 for F275W and -6.0 for F110W and F160W.}
\label{lumfun2}
\end{figure*}

\begin{figure*}[ht!]
\centering
\includegraphics[scale=0.65]{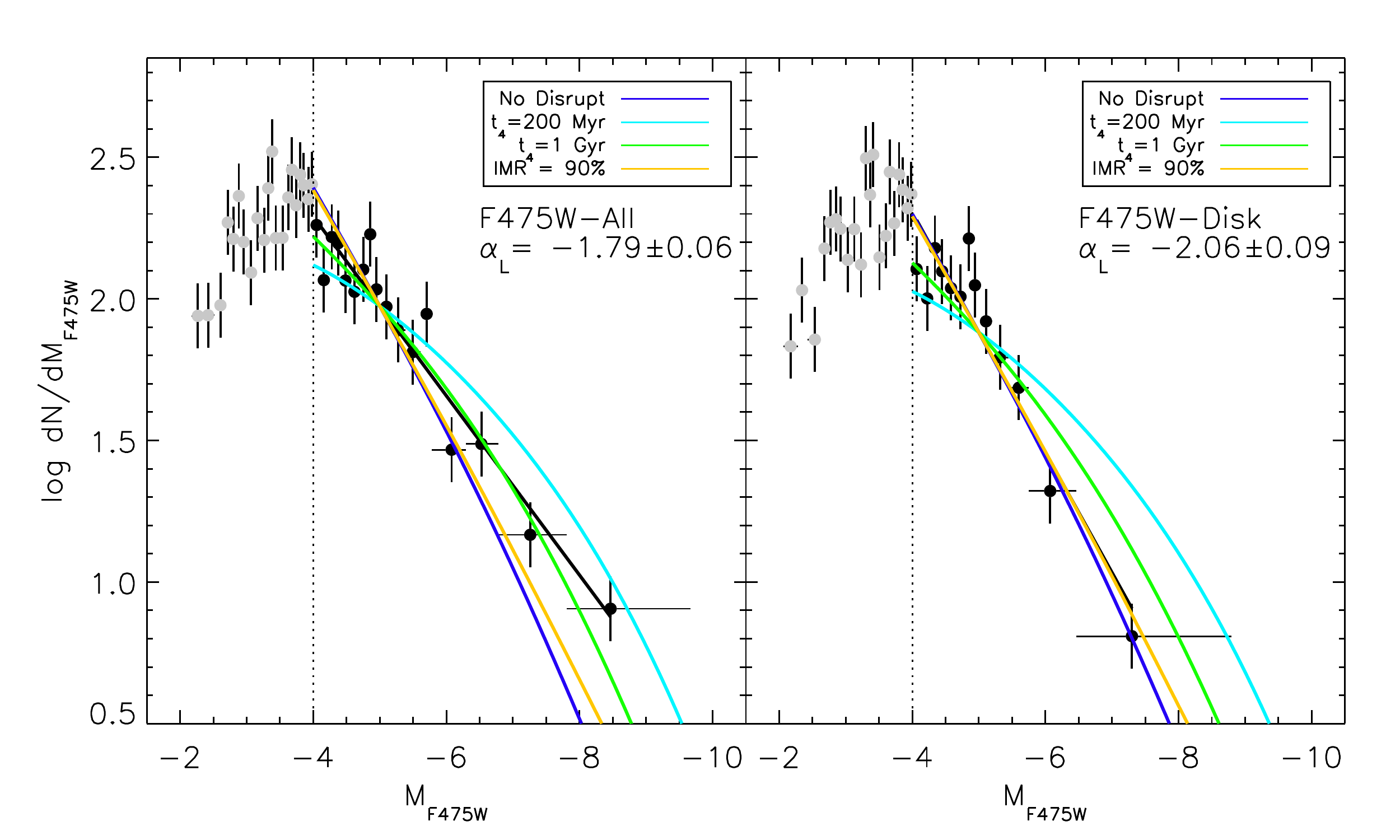}
\caption{F475W luminosity functions from Fig.~\ref{lumfun} with model luminosity functions from \citet{Larsen09} overplotted.  The left panel shows the luminosity function constructed using all Year 1 clusters, and the right panel shows the luminosity function constructed using a subset of Year 1 disk clusters that lie outside the bulge-dominated B01.  The outcomes of four different cluster dissolution models are compared: no disruption, two models of mass-dependent disruption as parameterized by \citet{Lamers05} for two dissolution timescales ($t_4$; characteristic destruction timescale for a $10^4$ \solmass\ cluster), and mass-independent disruption in which 90\% of clusters are destroyed within every logarithmic age interval.  Models are normalized to the data at M$_{F475W}=-5$.  Although the samples appear to prefer particular dissolution models, the effects that complex underlying cluster age and mass distributions impart on the luminosity functions do not allow us to draw any conclusions about cluster disruption at this time.}
\label{lumfunmod}
\end{figure*}

Luminosity functions are a basic, model independent measure of a stellar cluster population.  Up to this point, observations of spiral galaxies have provided largely consistent results, where luminosity functions are well fit by a power-law ($N \propto L^{-\alpha_{L}}$) with indices of about $-2$ or steeper \citep{Larsen02, Gieles10a, Chandar10-M83}.  In addition, there is growing evidence of steeper slopes at the brightest cluster luminosities \citep[M$_{V} < -9$;][]{Whitmore99,Gieles06a,Haas08}, suggesting the possibility of a Schechter-like cluster mass function with an exponential truncation at the high-mass end \citep{Larsen09}. 

These previous studies have generally focused on the bright end of the luminosity function, fitting clusters with luminosities greater than $\sim2 \times 10^{4}$ L$_\odot$ (equivalent to M$_{V} \lesssim -6$).  In contrast, the faint end of luminosity function has relatively poor constraints due to detection and catalog completeness difficulties encountered by previous extragalactic cluster studies.  However, due to our ability to identify low luminosity clusters in PHAT survey data, we can use the Year 1 cluster sample to probe the shape of the luminosity function down to limits only previously accessible in the Magellanic Clouds .  These new constraints are interesting because of the sensitivity of the faint end luminosity function to cluster disruption \citep{Larsen09}.

To assess the shape of the PHAT Year 1 luminosity functions in each of the six filter passbands, we consider two different samples of objects for fitting: all 601 clusters, and a subsample of 534 disk clusters that excludes objects from the bulge-dominated Brick 1.  The cluster population within Brick 1 is dominated by old, massive globular clusters associated with the galaxy's bulge component, in addition to the significantly brighter completeness limits in this region.  The disk sample represents a set of predominately younger objects that are likely more homogeneous in terms of formation and destruction history, simplifying the interpretation of its resulting luminosity functions.  We narrow the samples further to exclude clusters with uncertain photometry ($\sigma > 0.5$ mag), performing the quality cut individually for each passband.  The fraction of objects in each filter that meet this quality requirement is listed in Table~\ref{photdat}.  Finally, we convert aperture corrected, total magnitudes for the selected clusters to absolute magnitudes, correcting for the M31 distance modulus and Galactic foreground reddening.  We note that no correction for dust attenuation within M31 has been applied; we will derive these correction factors on an object-by-object basis as part of future age and mass fitting analysis (Fouesneau et al. 2012, in prep.).  

To characterize the power law slope of the luminosity function, we perform a linear fit to constrain the slope of log dN/dM (the logarithm of the number of clusters per magnitude) as a function of absolute magnitude.  For each passband, we bin the clusters using variable-sized magnitude bins to ensure fair weighting of the data, following the suggestion of \citet{DAgostino86} \citep[and more recently in][]{MaizAp05, Haas08} to group the data such that each bin represents an equal number of clusters, $N$.  We choose $N=15$, but find the results are insensitive to the particular number chosen.  We fit to datapoints brighter than absolute magnitude completeness limits: $-4.0$ for F275W, F336W, and F475W; $-5.0$ for F814W; $-6.0$ for F110W and F160W.  These limits are conservative; they are equivalent to $>$80\% completeness for all bricks, as found in Section~\ref{comp}.  We convert the resulting magnitude-based slopes to equivalent luminosity function slopes ($\alpha_{L}$) and report these as our primary results.  The luminosity functions and their associated power-law fits for the complete and disk-only cluster samples are presented in Figs.~\ref{lumfun} and \ref{lumfun2}; the values of the fitted slopes are listed in Table~\ref{lumfunfit}.  We separate the results into two groups to isolate the F275W, F110W, and F160W passbands, which carry the potential for larger systematic uncertainties due to the smaller fraction ($\sim$50-75\%) of clusters with well-determined photometry available at these wavelengths.  As such, we will focus our subsequent discussion on results from the F336W, F475W, and F814W passbands in Fig.~\ref{lumfun}.

Considering the complete sample of Year 1 clusters, we find luminosity function power law slopes in the F336W, F475W, and F814W passbands that are all flatter than $-2$.  While these measurements agree with the general trend of flatter slopes measured at fainter luminosities \citep[e.g., in NGC45 and M51;][]{Mora07, Haas08}, we suggest these flat slopes result from the inclusion of a relatively large number of luminous, evolved globular clusters associated with Brick 1 and included in the complete cluster sample.  We find that although clusters in Brick 1 make up $\sim$15\% of the complete sample by number, $\sim$50\% of sample members with $m_{F475W} < 19$ ($M_{F475W} \lesssim -5.5$) are Brick 1 clusters with spectroscopically derived ages of $>$10~Gyr \citep{Caldwell09,Caldwell11}.  Globular clusters are known to follow a Gaussian-like luminosity function shape with a peak magnitude around $M_{V}$ of $-7$ and $-8$, depending on factors such as metallicity and age \citep[e.g.,][]{Harris91, Barmby01a}.  The dissimilar luminosity function shape of the globular cluster subpopulation significantly influences the slope of the overall complete Year 1 luminosity function.

To determine the luminosity function shape that is characteristic of the dominant, young cluster population found in M31, we examine the fitting results for the disk-only sample of clusters, presented in the bottom row of Fig.~\ref{lumfun}.  We find uniformly steeper slopes ($\alpha_{L}<-1.9$) for the F336W, F475W, and F814W disk-only luminosity functions when compared to the complete Year 1 sample results.  In addition to the closer agreement between these slopes and the canonical $-2$ power law, we also observe a significant trend where the luminosity functions at bluer wavelengths are flatter than at redder wavelengths.  This behavior can be explained by wavelength and age-dependent cluster mass-to-light ratios \citep{Gieles10a}.

To visualize the sensitivity of the faint end luminosity function slope to differences in cluster dissolution, we compare our F475W luminosity functions to shapes predicted by four canonical dissolution models in Fig.~\ref{lumfunmod}.  The analytical luminosity functions were calculated by \citet{Larsen09} under the assumption of a constant cluster formation history, an underlying Schechter mass function (with cutoff of 2$\times 10^5$ \solmass), and continuous, non-stochastic sampling of the stellar initial mass function\footnote{We note that in terms of luminosity function modeling, we find little evidence for any noticeable differences when stochastic sampling of a cluster's stellar initial mass function is incorporated, validating the present model comparison.}.  In the case of mass-dependent cluster dissolution, \citet{Larsen09} finds that the slope of the luminosity function should flatten to $>-2$ at $M_{V} > -8$ or $-9$, while scenarios with mass-independent dissolution (and the case of zero dissolution) result in slopes that always remain steeper than $-2$.

The complete Year 1 cluster sample appears to be better described by a mass-dependent disruption model, while the steeper disk cluster luminosity function appears to agree with the shape of the mass-independent or the no disruption case.  However, due to the degenerate effects of underlying cluster age and mass distribution differences and variations in cluster disruption, such simple interpretation is not possible.  We opt to defer further analysis and interpretation until we obtain robust age, mass, and extinction determinations for individual clusters.  This crucial information will allow for the separation of the degenerate effects that influence the shape of the luminosity function and better constrain the characteristics of the underlying cluster population.

\subsection{Cluster Sizes} \label{size}

\begin{figure}[ht!]
\centering
\includegraphics[scale=0.56]{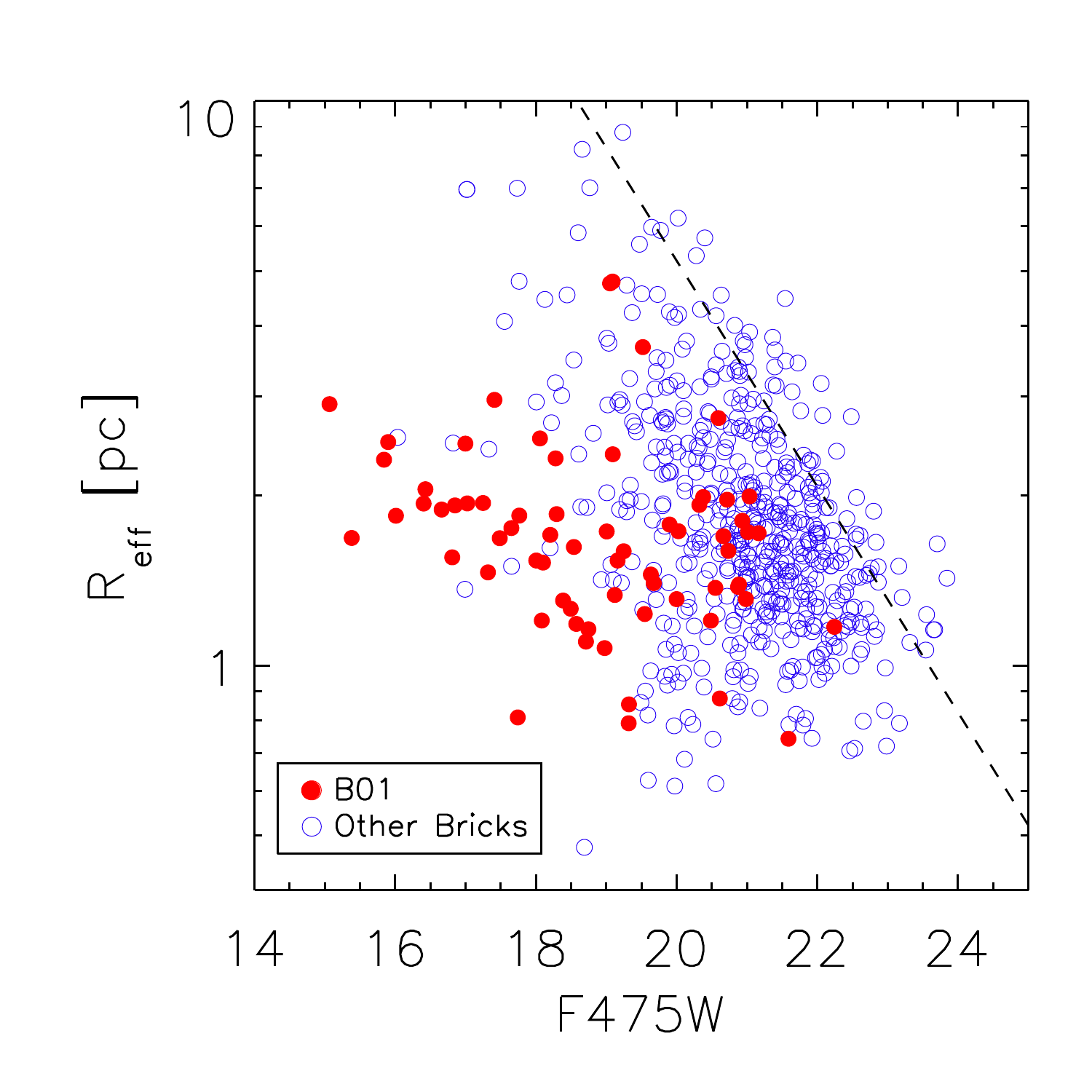}
\caption{$R_{eff}$ as a function of F475W magnitude.  The old, globular cluster dominated population in B01 (red solid circles) occupy a different region of parameter space than the younger disk clusters that dominate the remaining outer bricks (blue open circles).  The black dashed line denotes an approximate 50\% completeness limit, corresponding to a uniform surface brightness limit consistent with a F475W cutoff of 21.2 at an $R_{eff}$ of 3 pc.}
\label{magradius}
\end{figure}

\begin{figure*}[ht!]
\centering
\includegraphics[scale=0.65]{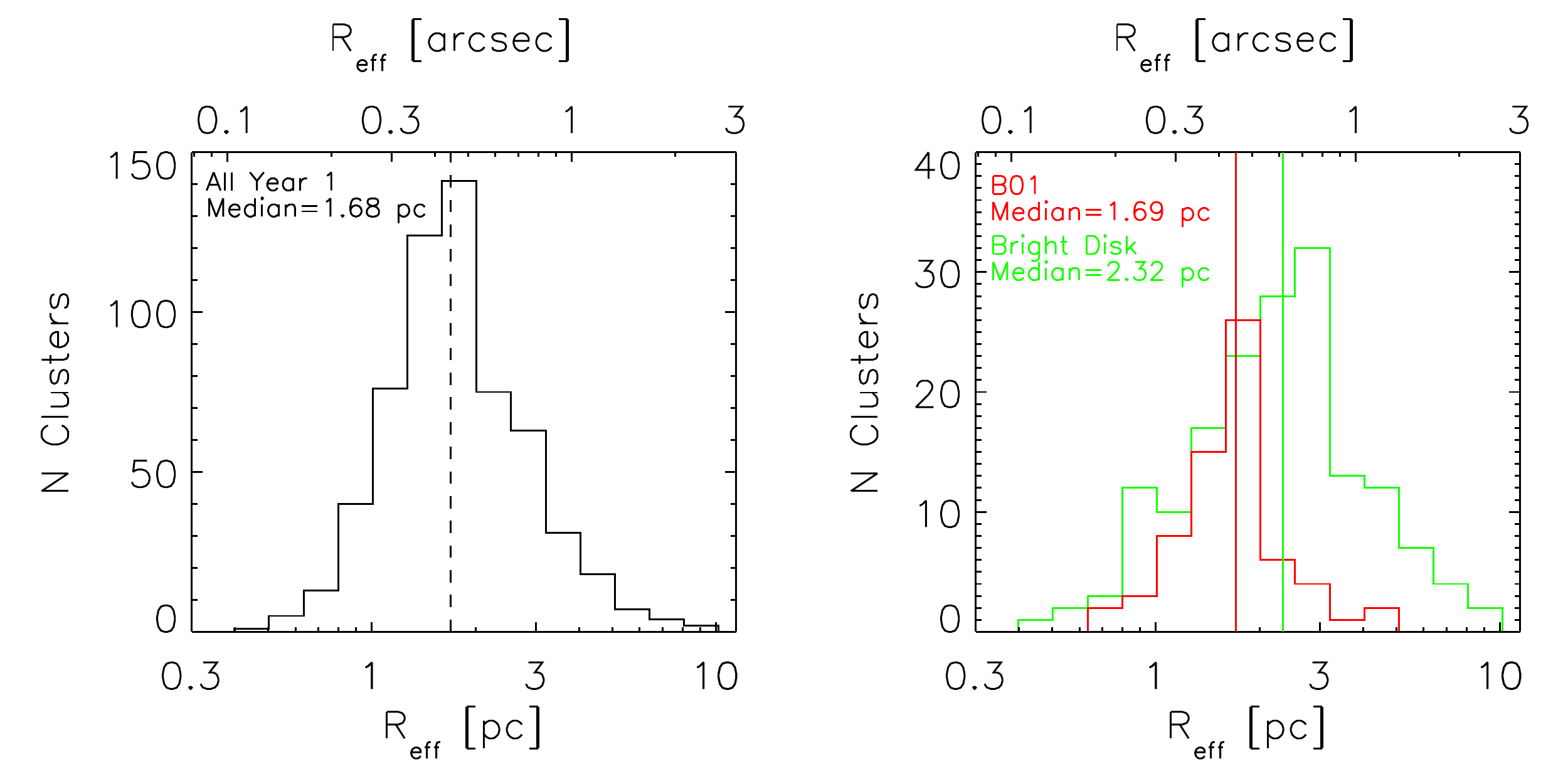}
\caption{Histograms of $R_{eff}$ derived from F475W data.  Left: Distribution of $R_{eff}$ for the full Year 1 cluster sample.  Median $R_{eff}$ is 1.68 pc (0.44$''$).  Right: Distribution of $R_{eff}$ for the B01 (red) and bright disk (green) subsamples.  Median $R_{eff}$ for the two subsamples are 1.69 pc (0.44$''$) and 2.32 pc (0.61$''$), respectively.}
\label{radhist}
\end{figure*}

Cluster size is a fundamental parameter that provides constraints on the dynamical state of stellar systems.  While many interesting correlations involving cluster sizes require the determination of other cluster parameters (e.g., mass-radius and age-radius relations), from sizes alone we can compare the overall properties of the cluster sample to those in other galaxies and look for interesting sub-populations of clusters within the sample.

We plot $R_{eff}$ as a function of F475W magnitude in Fig.~\ref{magradius}.  In this parameter space, the sample separates well into two groups: Brick 1 clusters that are predominately old ($\sim$10 Gyr), massive globular clusters associated with the galaxy bulge as discussed in Section~\ref{lum}, and clusters from the remaining bricks that are associated with Andromeda's star forming disk.  Overall, the Brick 1 clusters are more compact and luminous than the disk clusters, and show a trend in which $R_{eff}$ decreases with decreasing F475W luminosity.  Along with the cluster data, we plot a line of constant surface brightness anchored at F475W of 21.2 and a $R_{eff}$ of 3 pc, representing the 50\% catalog completeness limit as determined in Section~\ref{comp}.  The distribution of clusters with respect to this line suggests that, to first order, catalog completeness roughly follows a constant surface brightness cutoff.

The disk cluster sample includes objects with a broad range of ages and suffers from significant completeness truncations, making its magnitude-radius distribution difficult to interpret.  On the other hand, the Brick 1 clusters are similarly aged and are generally much brighter than the completeness limit.  This allows us to conclude that the moderate trend in which fainter Brick 1 clusters are more compact is real.  We do not currently know the physical origin of this trend, but plan to follow up using ages and metallicities from \citet{Caldwell11} and masses from \citet{Strader11}.

The left panel of Fig.~\ref{radhist} presents the principal result of our size analysis, which shows that the overall size distribution of the Year 1 PHAT cluster sample can be described approximately as a log-normal distribution, where the median $R_{eff}$ value is 1.68 pc (0.44$''$).  This median $R_{eff}$ is similar to, but smaller than values found in other galaxies.  In M83, \citet{Bastian11} find a overall median value of $\sim$2.4 pc for a sample of mostly young clusters, while \citet{McLaughlin05} find a median value of $\sim$3.3 pc for a sample of Milky Way globular clusters.  Additionally, \citet{Barmby07} find a median value of $\sim$2.5 pc for a sample of previously known M31 globular clusters.

We can explain the differences in median cluster sizes by considering how the composition of the Year 1 PHAT cluster sample differs from the M83, Milky Way, and existing M31 samples.  One possible reason for the median mismatch with the Milky Way and M31 globular cluster samples is that the Year 1 PHAT sample is a heterogeneous mix of mainly young clusters, while the  globular samples are uniformly much older.  However, an age-based explanation is ruled out because the old, globular-dominated Brick 1 subsample has a similar, but still smaller median $R_{eff}$ of 1.69 pc.  The $R_{eff}$ distribution for the Brick 1 subsample is shown in the right panel of Fig.~\ref{radhist}.

An explanation for the median differences becomes clear when we consider the galactocentric radii of the existing globular cluster samples.  A correlation between $R_{eff}$ and galactocentric radius has been show to exist in many galaxies such that inner clusters are more compact \citep{Barmby07}.  When we limit the Milky Way and previous M31 cluster samples to include only objects with galactocentric radii less than 3 kpc, which roughly matches the radial extend of the Brick 1 sample, the median $R_{eff}$ for these two samples drops to $\sim$2.0 pc.  These lower medians are in better agreement with the Brick 1 subsample, easing the tension between the values derived for the two globular cluster samples and for the PHAT clusters.

To explain the median $R_{eff}$ difference between the young M83 cluster sample and the Year 1 PHAT sample, instead of age and galactic position, we consider how completeness limits influence the shape of the $R_{eff}$ distribution.  In terms of numbers, the Year 1 cluster sample is dominated by compact, low luminosity objects that lie just above our detection threshold.  At these faint magnitudes, surface brightness limits prevent us from detecting clusters across the full range of possible $R_{eff}$ values.  As a result, the overall cluster sample is biased towards compact objects, driving the median towards smaller values.  However, if we only consider clusters with F475W $< 20.5$, the biasing effects of surface brightness limits are greatly reduced.  The $R_{eff}$ distribution for this luminous subsample is shown in the right panel of Fig.~\ref{radhist}.  Without the faint, compact portion of the sample, the median cluster size rises to 2.32 pc, which agrees well with the observed value obtained for clusters in M83.  In conclusion, cluster sizes for the Year 1 sample are consistent with results derived by previous works in similar spiral galaxies once sample completeness and composition are taken into account.

\subsection{Objects of Interest} \label{obj}

\subsubsection{Red Supergiant Clusters} \label{rsg}

\begin{figure*}
\centering
\includegraphics[scale=0.26]{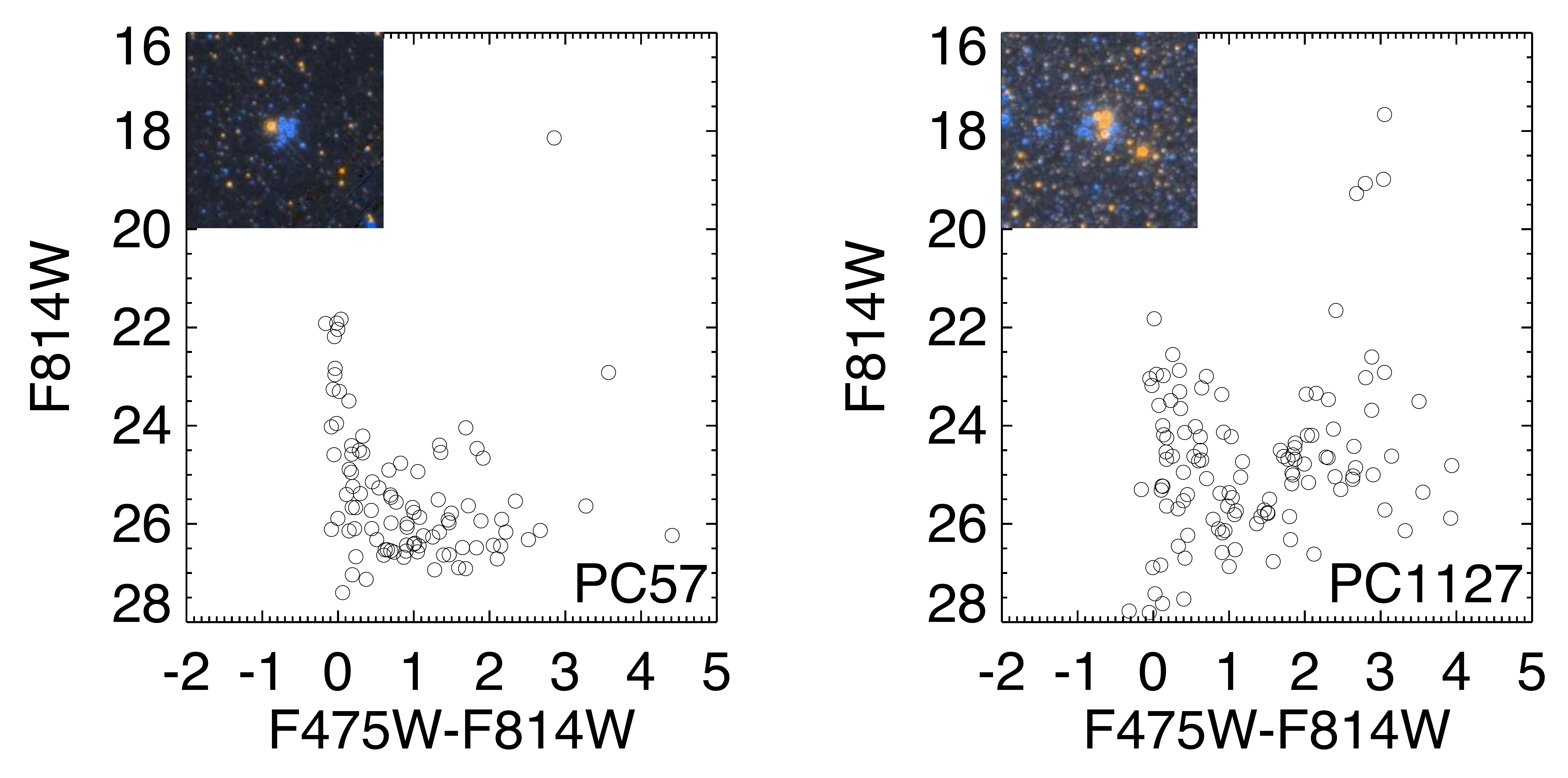}
\caption{Resolved star color-magnitude diagrams and color optical (F475W+F814W) image cutouts for two candidate red supergiant clusters.  These clusters have unusually red F475W-F814W colors due to the presence of luminous evolved stars that strongly bias integrated light measurements.}
\label{rsgfig}
\end{figure*}

We identified a subsample of objects with unusually red F475W-F814W colors in Fig.~\ref{colorcolor}.  Within the selection box, we identify $\sim$15 clusters that show extreme integrated colors. Upon inspection, we find that many of these clusters appear to host luminous red supergiant (RSG) stars, which explain their anomalous integrated colors.  Two examples of this class of cluster, PC57 and PC1127, are shown in Fig.~\ref{rsgfig}.

Red supergiant stars are massive (8 to 25 \solmass) stars that have left the main sequence and are traversing the top of the CMD during the helium burning phase of stellar evolution.  Individual luminous RSG stars have the ability to bias the integrated colors of low mass ($<10^4$ \solmass) clusters because stochastic sampling of the stellar mass function in low mass systems can cause the small number of evolved stars to vary by factors of a few (low mass clusters typically host between 0 and 4 evolved stars).  This variation creates wild fluctuations in the total luminosity and integrated color of the cluster, which depend on the particular number of evolved stars progressing through this relatively short evolutionary phase.

The existence of these color outliers confirms that stellar population modeling of low mass clusters must account for stochastic sampling of the stellar initial mass function to obtain accurate age and mass determinations \citep[see e.g.,][]{Fouesneau10}.  Models that assume a fully sampled stellar initial mass function cannot reproduce objects with integrated colors as red as those shown by RSG clusters.  Our future cluster analysis will benefit from the use of cluster models that account for stochastic fluctuations in integrated light caused by the RSGs.

More importantly, our ability to resolve individual stars in clusters provides us a number of important benefits with respect to the RSG clusters.  Resolved star photometry enables us to use CMD fitting techniques to obtain cluster parameter determinations, allowing us to avoid the use of biased integrated measurements all together.  In addition, once we obtain age, mass, and attenuation information for the clusters, we can use this information to tag individual RSGs and use these constraints to improve calibration of late stage stellar evolution for massive stars at high metallicity.

\subsubsection{Massive Clusters} \label{massive}

Globular clusters, with old characteristic ages ($\sim$10 Gyr) and large masses ($>10^5$ \solmass), have long been a target of study in M31.  Although the Year 1 cluster sample is numerically dominated by low mass clusters, it contains many massive clusters as well.  While we identified few new globular clusters as part of the Year 1 search, our high spatial resolution imaging has enabled us to confirm or eliminate a large number of globular cluster candidates from existing catalogs.  In addition to the 63 out of 71 highly-ranked clusters we confirm from the RBC, we affirm 12 and reject 19 possible candidates.  The PHAT survey also enables detailed analysis of these objects by means of resolved star photometry for cluster members.  We plan to place better constraints on red giant branch (RGB) and horizontal branch (HB) morphology as a function of metallicity through these detailed studies.

In addition to the old globular clusters, we have also newly identified a number of intermediate mass ($\sim10^4$ \solmass), intermediate age ($\sim$1 to 3 Gyr) clusters.  These objects are interesting targets for study because of the relative rarity in the Milky Way \citep{Friel95}, where most similarly aged Galactic clusters are less massive.  While a small number of these intermediate mass, intermediate age clusters are already known to exist in M31 \citep{Caldwell09,Caldwell11}, increasing the sample size for this class of object should help to better understand their origin and evolution.  Investigation of these objects will complement the study of younger ($<$1 Gyr) massive clusters by \citet{FusiPecci05}, \citet{Caldwell09}, and \citet{Perina10}.

\begin{figure*}
\centering
\includegraphics[scale=0.3]{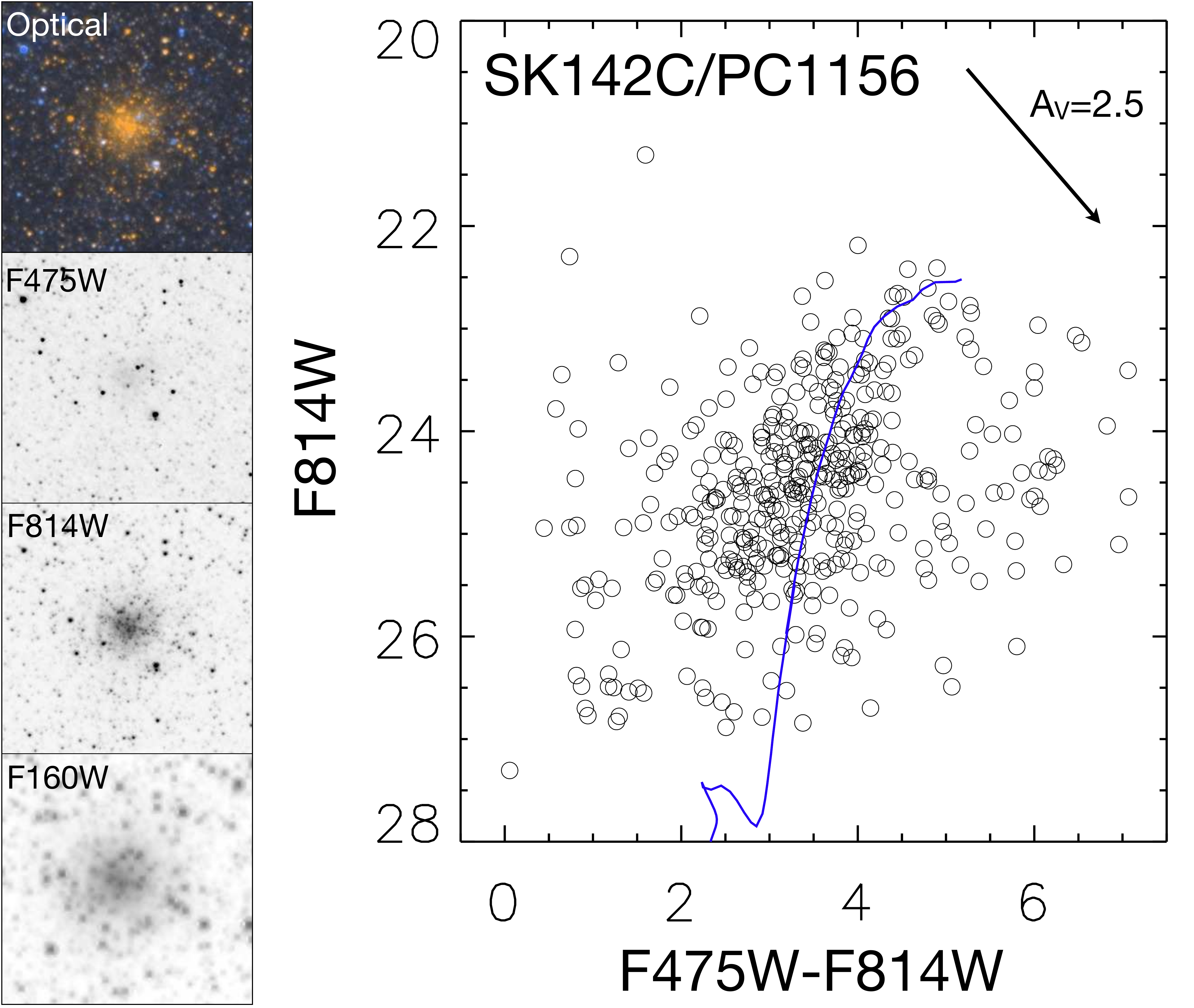}
\caption{Left: Image cutouts showing highly-reddened, intermediate age cluster SK142C (PC1156).  Right: The resolved star color-magnitude diagram for SK142C shows a reddened RGB, indicative of an intermediate age population.  We overplot a Padova stellar isochrone with a log(Age) of 9.35 ($\sim$2 Gyr), $A_V$ of 2.5 mag, and solar metallicity to show that the cluster is consistent with these parameter values.}
\label{sk142cfig}
\end{figure*}

The most remarkable massive cluster in the Year 1 catalog is the cluster SK142C (PC1156).  This object is the sample's most discrepant color outlier, appearing at the redward extreme of all cluster photometric measurements, with a F475W-F814W color of $\sim$3.5.  This cluster appears to be an intermediate mass, intermediate age, highly-reddened cluster.  Along with cutout images showing the cluster's heavily reddened appearance, we plot the cluster's resolved star CMD in Fig.~\ref{sk142cfig}.  The CMD shows an extended, highly-reddened RGB, indicative of an intermediate cluster age.  We perform a visual comparison to solar metallicity Padova isochrones \citep{Girardi10} and find that the CMD is consistent with an age of $\sim$1--3 Gyr and $A_{V}$ of 2.25--2.75 mag, assuming large uncertainties associated with visual ``chi-by-eye'' isochrone fitting.  Using this combination of age and reddening, we use the cluster's F160W luminosity to estimate a cluster mass of $\sim 1.5 \times 10^5$ \solmass.  Due to its massive nature and its somewhat intermediate age, this cluster is a rare and interesting object akin to newly discovered objects in the Milky Way \citep{Figer06,Davies11,Davies11a}.

\section{Summary \& Future Work} \label{conclusion}

In this paper, we presented the first installment of the PHAT stellar cluster catalog.  We introduced the Year 1 cluster sample, consisting of 601 clusters identified through a visual search of high spatial resolution HST imaging.  The PHAT cluster sample significantly increases the number of clusters known in M31; the catalog presented here represents more than a factor of four increase in the number of known clusters within the current survey area.  We presented a basic assessment of the cluster sample, including positional, size, and photometric information.

We have shown that the PHAT cluster sample hosts a large range in ages and masses.  This wide-ranging sample provides the opportunity for a top-to-bottom study of stellar cluster evolution processes.  The PHAT sample includes low-luminosity objects missed in studies of distant external galaxies, while covering an uninterrupted range of cluster masses, unlike Milky Way cluster samples.

When the survey is complete, the PHAT cluster catalog will be among the largest and most comprehensive surveys of star clusters in any galaxy.  This work presents results derived from the first 25\% of the survey data; we estimate that the final sample will include $\sim$2500 clusters.  Over the duration of the survey, we plan to periodically publish updates to the catalog to include new clusters and to revise object classifications as we gather additional information about the sample.  Age and mass determination analysis for the Year 1 sample is currently underway (Fouesneau et al. 2012, in prep.), which will provide the means to explore a host of different topics, including the cluster mass function and cluster dissolution.  The Year 1 cluster sample will also be the basis for analysis of structural parameters, resolved star content, and integrated spectroscopy, but we also look forward to future studies enabled by the complete, four year PHAT cluster catalog.


\acknowledgements
{The authors wish to acknowledge the collective efforts of the entire PHAT team in this project.   Also, the authors thank the anonymous referee for a prompt and useful report.  This research made extensive use of NASA's Astrophysics Data System Bibliographic Services.  Support for this work was provided by NASA through grant number HST-GO-12055 from the Space Telescope Science Institute, which is operated by AURA, Inc., under NASA contract NAS5-26555.  D.A.G. acknowledges financial support from the German Aerospace Center (DLR) and the German Research Foundation (DFG) through grants 50~OR~0908 and GO~1659/3-1, respectively.}

{\it Facilities:} \facility{HST (ACS, WFC3)}.


\appendix

\section{Background Galaxy Catalog} \label{bckgal}

As introduced in Section~\ref{byeye}, we present a catalog of objects visually identified as background galaxies in Table~\ref{galcat}.  Along with positional information, we present a rough assessment of the galaxy size.

\section{Commentary on Existing Cluster Catalogs} \label{revise}

We performed a comparison of object classifications from existing cluster catalogs to those in the PHAT Year 1 cluster catalog in Section~\ref{oldcat}.  Here we provide comments on individual catalog comparisons, including detailed classification statistics and object-by-object classification revisions.  We discuss individual classification statistics for the RBC \citep{Galleti04}, \citet{Kim07}, \citet{Caldwell09, Caldwell11}, \citet{Peacock10}, and three of the four HKC studies (\citealt{Krienke07,Hodge09,Hodge10}; no objects from \citealt{Krienke08} lie within the Year 1 PHAT footprint).  Classification results broken down by object type are presented in Table \ref{oldcat-summary}.

\textit{The Revised Bologna Catalog} --- Our search for RBC objects that lie within the Year 1 PHAT imaging footprint produces a catalog of 187 objects.  These objects include 71 clusters, 36 cluster candidates, and 80 non-cluster objects.  After reanalyzing these classifications using the by-eye ranking results and following up on other objects that were not selected as PHAT candidates, our revised object classifications show excellent agreement with the original RBC results.  Individual object revisions are provided in Table \ref{revise-rbc}, but the bulk of the changes are due to our ability to determine the nature of the RBC's candidate objects (RBC Flag = 2).  Of the 36 candidates, we confirm 12 objects as clusters in the PHAT catalog, reject 23 candidates (including two originally controversial cases, one duplicate entry, and one object not found in PHAT imaging), and transfer the final candidate to the PHAT possible cluster sample.  In terms of catalog accuracy, only five ``confirmed'' clusters (RBC Flag = 1) were confidently revised to non-cluster objects, four of which were \citet{Kim07} objects (see additional discussion below).  Overall, the high level of consistency between the two catalogs gives us great confidence in galaxy-wide accuracy of the RBC cluster sample.

\textit{\citet{Kim07}} --- The catalog content of \citet{Kim07} is also accounted for in the results of the RBC cross-comparison, but the resulting cluster confirmation and rejection statistics from this work is worth individual mention.  As pointed out in \citet{Caldwell09} and \citet{Peacock10}, the \citet{Kim07} catalog has a high level of contamination, namely from misclassified stellar sources.  Considering all three cluster quality categories from \citet{Kim07} together, we find 38 objects from this catalog that fall within the Year 1 PHAT footprint.  Of these, only 37\% of those objects appear in the PHAT catalog as confirmed or possible clusters.  Even when taking the quality rankings into consideration, we find that (50\%, 27\%, 38\%) of the (A, B, C) objects appear in the catalogs from our present work.

\textit{\citet{Caldwell09, Caldwell11}} --- These two papers revised classifications from v3.4 of the RBC (many of which were adopted in v4.0) using low-resolution spectra, ground-based imaging from the Local Group Galaxy Survey \citep[LGGS;][]{Massey06}, and HST-based images.  As this catalog has been incorporated into the latest version of the RBC, discussion of individual object classifications are included in Table \ref{revise-rbc}.  When comparing our current PHAT classification with those from \citet{Caldwell09,Caldwell11}, we find good agreement between the two works.  We revise classifications for 6 clusters out of 183 that fall within the Year 1 footprint.

\textit{\citet{Peacock10}} --- The clusters considered by \citet{Peacock10} are included in the RBC comparison discussed above, and classifications for each of these objects is already provided in Table \ref{revise-rbc}.  Many of the reclassifications performed in \citet{Peacock10}, which was based on v3.5 of the RBC, have proven to be accurate (including the stellar classification of numerous \citealt{Kim07} candidates).  Overall, we find good consistency between the \citet{Peacock10} catalog and our PHAT classifications.

\textit{Hodge-Krienke Catalogs} --- Similar data and methodology (including common authors in the case of P.H.) leads to strong consistency between the HKC and the work presented here.  We find 54 clusters from the HKC that match entries in the PHAT cluster catalog and 8 additional objects that match possible clusters, out of a total of 75 objects that lie within the boundaries of the PHAT Year 1 footprint.  Unlike in the case of ground-based catalogs, the 13 HKC objects not recovered as part of the PHAT cluster search mostly fall into the category of asterisms.  In other words, these objects were interpreted by the PHAT search team as uncertain or unlikely cluster candidates, and omitted from the final catalog presented in this work.  This subset of objects shows that for the least luminous, least massive clusters, defining the difference between an object that is potentially a bound stellar system and simply a chance collection of stars seen in projection becomes a difficult and subjective task.  Table \ref{revise-hodge} presents proposed revisions to the HKC.  In closing, we note that we do not compare to cluster identifications presented in \citet{Williams01} because classifications in this work were superseded by the HKC.

\section{Comparison to Existing Cluster Photometry} \label{photcompare}

We provide a comparison between PHAT cluster photometry and the results of \citet{Peacock10}, \citet{Fan10}, \citet{Hodge09}, and \citet{Hodge10}.  We compare photometry for these selected studies because they provide magnitudes derived in similar photometric passbands as part of a uniform analysis.  We choose not to compare to photometry provided by the Revised Bologna Catalog due to the heterogenous nature of their unified photometric measurements \citep[see][for details]{Galleti04}.  We note that both \citet{Peacock10} and \citet{Fan10} find good agreement between photometric results and that of \citet{Barmby00}, upon which the photometry of the Revised Bologna Catalog is based.

\begin{figure*}
\centering
\includegraphics[scale=0.65]{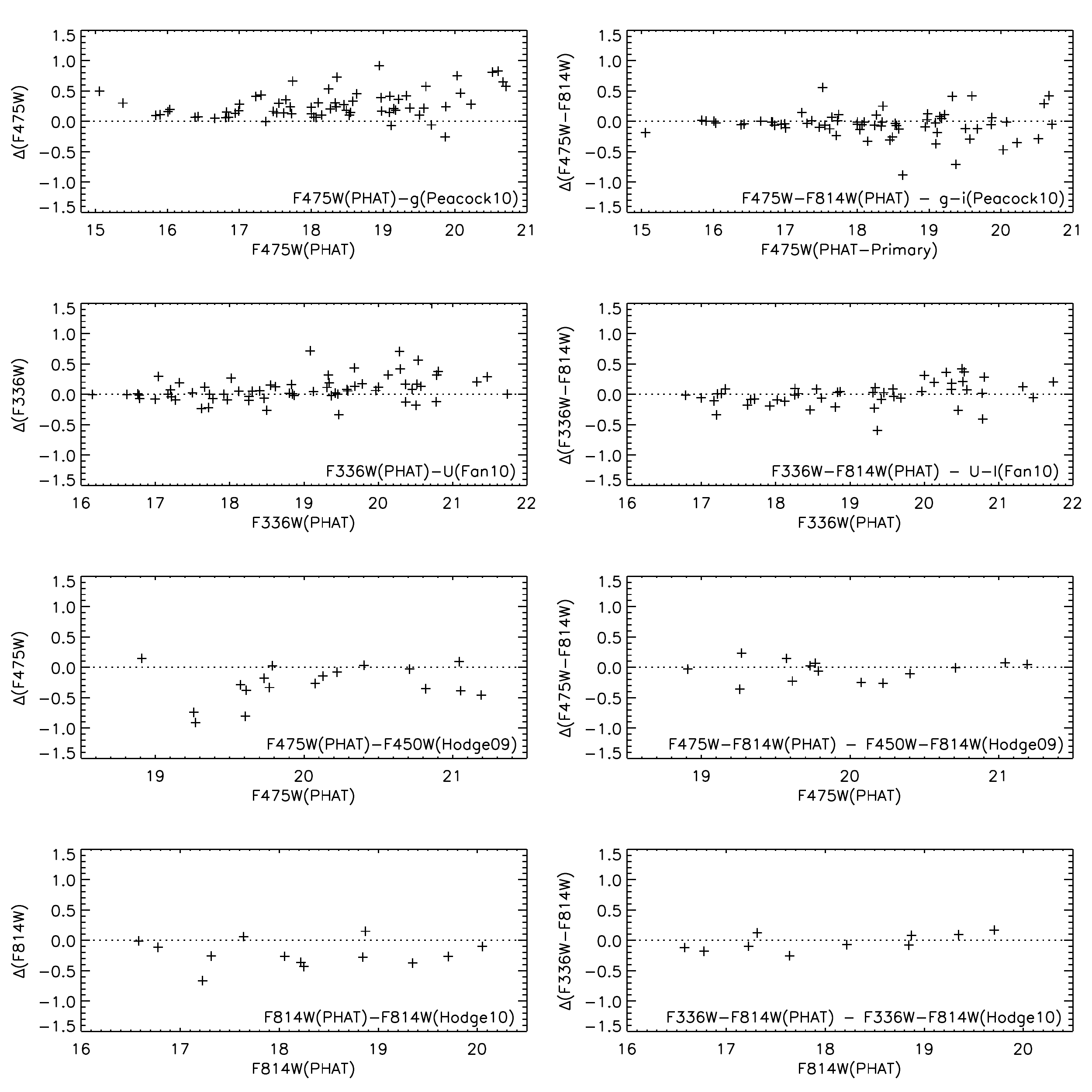}
\caption{Comparison of PHAT photometric results with those of \citet{Peacock10} (top row), \citet{Fan10} (second row), \citet{Hodge09} (third row), and \citet{Hodge10} (bottom row).}
\label{photcomp}
\end{figure*}

\begin{figure}
\centering
\includegraphics[scale=0.7]{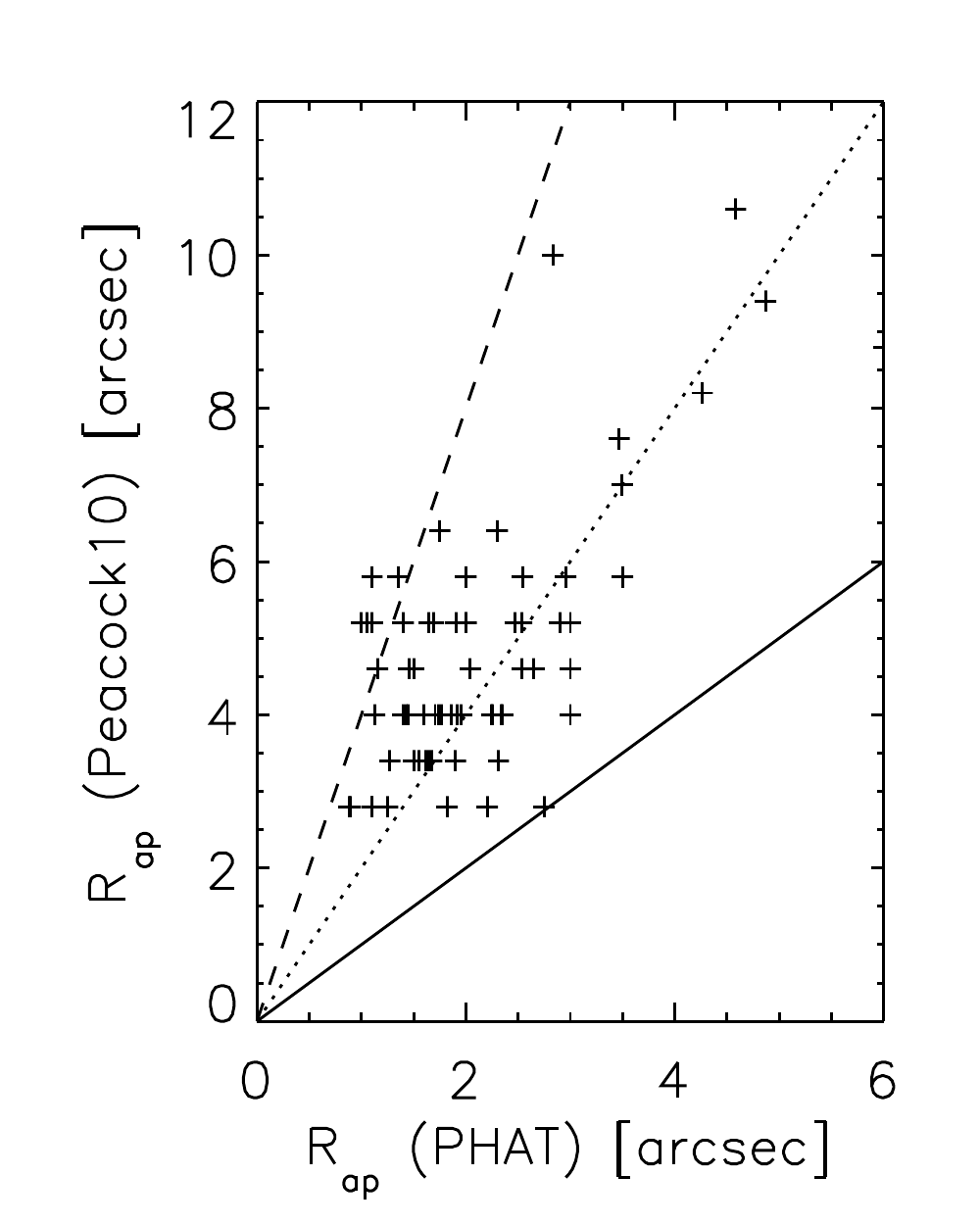}
\caption{Comparison of photometric aperture radii used in \citet{Peacock10} and in this work.  The solid line represents a 1-to-1 relation, the dotted line a 2-to-1 relation, and the dashed line a 4-to-1 relation.}
\label{comp_p10ap}
\end{figure}

The optical aperture photometry of \citet{Peacock10} was derived from SDSS $ugriz$ imaging.  We compare photometric results for 77 clusters found in both datasets.  We calculate transformations to convert from AB-based $g$ and $i$-band photometry to Vega-based F475W and F814W magnitudes using transformations from \citet{Girardi08}, assuming a single-age, 1 Gyr solar metallicity population that represents a median value for the age-dependent transformation:
\begin{equation}
\label{trans475g}
F475W = g_{AB} + (0.06 \pm 0.02) ,
\end{equation}
\begin{equation}
\label{trans814i}
F814W = i_{AB} - (0.54 \pm 0.10) .
\end{equation}
The resulting photometric comparison is presented in the top row of Fig.~\ref{photcomp}.  We find that the PHAT magnitudes are fainter than those of \citet{Peacock10} by $\sim$0.25 mag.  This offset can be explained by the difference in aperture sizes, as seen in Fig.~\ref{comp_p10ap}.  The \citet{Peacock10} apertures have radii that are twice as large on average, and up to four times as large as those used in this work.  In the case of the brightest comparison cluster, B127-G185 (PC1425), the size of the aperture (10$''$) was large enough to include the nearby cluster NB89 (PC1426), explaining the $\sim$0.5 mag difference for this object.  In a comparison of F475W-F814W versus ($g-i$) optical colors, the agreement is quite good for brighter clusters (F475W $<$17) with increasing scatter and a redward bias of \citet{Peacock10} colors for fainter clusters.  These differences are likely caused by field contamination within the large \citet{Peacock10} apertures, and by blending in their low-resolution (1-2$''$ seeing) ground-based imaging.  In summary, we find systematic differences between the photometry derived in this work and that of \citet{Peacock10}, but the overall agreement is adequate and remaining differences can be readily explained by the effects of image resolution and aperture size differences.

In another recent compilation of ground-based aperture photometry, \citet{Fan10} derived Johnson-Kron-Cousins $UBVRI$ magnitudes from LGGS imaging \citep{Massey06}.  The Vega-based PHAT F336W and F814W magnitudes are nearly equivalent to the $U$ and $I$ magnitudes from \citet{Fan10}, allowing us to directly compare photometric results for 68 clusters common to both samples.  We present the results in the second row of Fig.~\ref{photcomp}, showing good agreement between the two datasets.  While \citet{Fan10} magnitudes appear slightly brighter for faint clusters, the overall agreement is improved with respect to the \citet{Peacock10} results.  One explanation for the improvement between ground-based and HST-based results could trace back to the superior image quality of the LGGS data when compared to the SDSS imaging of \citet{Peacock10}.  The average seeing for the LGGS data is $\sim$1$''$, while varying between 1-2$''$ for the SDSS imaging.  This results in the ability for \citet{Fan10} to reduce their aperture sizes, such that they are only a factor of $\sim$1.5 bigger than those used by PHAT.

The photometry of \citet{Hodge09} and \citet{Hodge10} allows us to compare two sets of HST-based photometric results.  Out of the four M31 cluster catalog papers that compose the HKC, we choose two where the tabulated photometry is provided in native, Vega-based HST filter magnitudes.  Although the data used in these studies are similar to those used in this work, there are fundamental differences in the photometry techniques employed.  As opposed to our growth curve method of aperture size determination, the HKC adopt an isophotal aperture determination, such that apertures extend out to a chosen surface brightness limit.  Further, the methods of sky background determination differ.  Our sky levels are determined using local sky background estimates measured in annular rings around the photometric aperture, while the HKC uses samplings of the sky background taken across the HST image.  Due to these differences in methodology and the lack of any correction to account for total cluster light, we expect the comparison to show some scatter and a systematic offset such that the PHAT photometry is brighter.

For the \citet{Hodge09} comparison, we use 18 objects common to both datasets and present the results in the third row of Fig.~\ref{photcomp}.  We compare our F475W photometry to magnitudes measured in a similar F450W passband, using a transformation derived in the same way as Eqs.~\ref{trans475g} and \ref{trans814i}: 
\begin{equation}
F475W = F450W - (0.11 \pm 0.10) .
\end{equation}
There are several significant ($\Delta >$0.5 mag) outliers in the F475W magnitude comparison, in which the photometry from this work is brighter by up to $\sim$1 mag.  To check the validity of our measurements, we performed photometry for the three most discrepant outliers on the WFPC2 images used by \citet{Hodge09}, employing photometry procedures used in this work.  We found resulting magnitudes that agree within $\sim$0.1 mag of our original PHAT photometry values.  This result indicates that the magnitude offsets are likely due to differences in photometric technique, specifically related to sky background determination.

The bottom row of Fig.~\ref{photcomp} presents the comparison of \citet{Hodge10} photometry for 13 common objects.  We find better overall agreement with this dataset when compared to the \citet{Hodge09} results.  We compare F814W magnitudes (no filter transformations required) and F336W-F814W colors, and find a scatter of $\sim$0.2 mag between the two datasets.  The comparison of photometry between our current PHAT study and the work of \citet{Hodge09, Hodge10} shows good overall consistency.  Systematic biases and a small number of significant outliers likely stem from differences in measurement technique.


\bibliographystyle{apj}
\bibliography{clusterlit}

\newpage


\begin{deluxetable*}{lccccccccr}
\tabletypesize{\tiny}
\setlength{\tabcolsep}{0.05in}
\tablecaption{PHAT Year 1 Cluster Catalog \label{tbl1}}
\tablewidth{0pt}

\tablehead{
\colhead{PC ID} & \colhead{RA (J2000)} & \colhead{Dec (J2000)} & \colhead{F275W}  & \colhead{$\sigma$} & \colhead{F336W}  & \colhead{$\sigma$} & \colhead{F475W} & \colhead{$\sigma$} & \colhead{ApCor\tablenotemark{a}} \\
\colhead{$S_{by-eye}$} & \colhead{$R_{ap}$ ($''$)} & \colhead{$R_{eff}$ ($''$)} & \colhead{F814W}  & \colhead{$\sigma$} & \colhead{F110W}  & \colhead{$\sigma$} & \colhead{F160W} & \colhead{$\sigma$}  & \colhead{Alternate Name}
}

\startdata
   1 & 11.638274 & 42.193887 & 17.54 & 0.02 & 17.68 & 0.03 & 18.82 & 0.02 & -0.12 \\
1.00 & 1.57 & 0.68 & 18.34 & 0.04 & 17.95 & 0.07 & 17.73 & 0.12 & Hodge10-85 \\ \hline
   2 & 11.637139 & 42.209936 & 15.60 & 0.02 & 15.91 & 0.02 & 17.33 & 0.01 & -0.00 \\
1.43 & 2.51 & 0.64 & 17.23 & 0.04 & 17.39 & 0.19 & 17.30 & 0.35 & Hodge10-84 \\ \hline
  20 & 11.630550 & 42.200631 & \nodata & \nodata & 22.55 & 0.25 & 21.97 & 0.63 & -0.13 \\
1.00 & 1.10 & 0.48 & 19.83 & 0.07 & 19.16 & 0.23 & 18.63 & 0.39 & \nodata \\ \hline
  21 & 11.631591 & 42.199991 & 19.60 & 0.02 & 19.82 & 0.01 & 20.89 & 0.10 &  -0.21 \\
1.57 & 1.00 & 0.52 & 20.98 & 0.28 & 22.88 & 1.80 & \nodata & \nodata & \nodata \\ \hline
  22 & 11.630849 & 42.201656 & 23.38 & 0.36 & 22.52 & 0.11 & 22.22 & 0.06 & -0.31 \\
1.71 & 0.75 & 0.46 & 21.39 & 0.67 & 22.47 & 1.72 & \nodata & \nodata & \nodata 
\enddata

\tablecomments{Table \ref{tbl1} is published in its entirety in the electronic edition of the {\it Astrophysical Journal}.  A portion is shown here for guidance regarding its form and content.  The sample presented here consists of objects classified as clusters, with $S_{by-eye} < 2.0$.}
\tablenotetext{a}{Aperture Corrections are provided such that $m_{Total} = m_{Aperture} + ApCor$.}

\end{deluxetable*}


\begin{deluxetable*}{lccccccccr}
\tabletypesize{\tiny}
\setlength{\tabcolsep}{0.05in}
\tablecaption{PHAT Year 1 Possible Cluster Catalog \label{tbl2}}
\tablewidth{0pt}

\tablehead{
\colhead{PC ID} & \colhead{RA (J2000)} & \colhead{Dec (J2000)} & \colhead{F275W}  & \colhead{$\sigma$} & \colhead{F336W}  & \colhead{$\sigma$} & \colhead{F475W} & \colhead{$\sigma$} & \colhead{ApCor\tablenotemark{a}} \\
\colhead{$S_{by-eye}$} & \colhead{$R_{ap}$ ($''$)} & \colhead{$R_{eff}$ ($''$)} & \colhead{F814W}  & \colhead{$\sigma$} & \colhead{F110W}  & \colhead{$\sigma$} & \colhead{F160W} & \colhead{$\sigma$}  & \colhead{Alternate Name}
}

\startdata
   4 & 11.664001 & 42.192391 & 19.37 & 0.06 & 19.43 & 0.05 & 20.38 & 0.02 & -0.09 \\
   2.43 & 2.10 & 0.83 & 19.80 & 0.14 & 19.42 & 0.22 & 18.63 & 0.10 & \nodata \\ \hline
   6 & 11.668817 & 42.197452 & 21.50 & 1.01 & 20.93 & 0.13 & 21.86 & 0.29 & -0.01 \\
   2.29 & 1.13 & 0.32 & 21.97 & 1.72 & 21.57 & 0.69 & 21.71 & 3.33 & \nodata \\ \hline
   7 & 11.675393 & 42.193555 &  \nodata & \nodata & 25.59 & 2.19 & 22.63 & 0.14 & -0.07 \\
   2.29 & 0.80 & 0.29 & 20.96 & 0.13 & 20.38 & 0.11 & 19.73 & 0.08 & \nodata \\ \hline
  11 & 11.648244 & 42.227958 & 25.42 & 1.35 & 23.98 & 0.24 & 22.56 & 0.15 & -0.08 \\
  2.43 & 1.00 & 0.38 & 20.12 & 0.12 & 18.77 & 0.03 & 17.82 & 0.07 & \nodata \\ \hline
  12 & 11.661360 & 42.190571 & 21.39 & 0.16 & 21.02 & 0.05 & 21.38 & 0.04 & -0.02 \\
  2.00 & 1.05 & 0.31 & 20.74 & 0.06 & 20.82 & 0.13 & 21.39 & 1.90 & \nodata
\enddata

\tablecomments{Table \ref{tbl2} is published in its entirety in the 
electronic edition of the {\it Astrophysical Journal}.  A portion is shown here for guidance regarding its form and content.  The sample presented here consists of objects classified as possible clusters, with $2.0 \leq S_{by-eye} < 2.5$.}
\tablenotetext{a}{Aperture Corrections are provided such that $m_{Total} = m_{Aperture} + ApCor$.}

\end{deluxetable*}


\begin{deluxetable*}{lc}
\tablecaption{Photometric Zeropoints \label{zeropoints}}
\tablehead{ \colhead{Passband} & \colhead{$Vegamag$ Zeropoint} }
\tablewidth{0pt}
\startdata
F275W & 22.65 \\
F336W & 23.46 \\
F475W & 26.16 \\
F814W & 25.52 \\
F110W & 26.07 \\
F160W & 24.70 
\enddata
\end{deluxetable*}


\begin{deluxetable*}{lcc}
\tablecaption{Passband Photometric Quality Comparison for Cluster Sample\label{photdat}}
\tablehead{\colhead{Passband} & \colhead{N(Valid Measurements)} & \colhead{N(Well-determined Measurements)}}
\tablewidth{0pt}
\startdata
F275W & 552 (91.8\%) &  447 (74.4\%) \\
F336W & 590 (98.2\%) &  566 (94.2\%) \\
F475W & 600 (99.8\%) &  597 (99.3\%) \\
F814W & 593 (98.7\%) &  514 (85.5\%) \\
F110W & 518 (86.2\%) &  358 (59.6\%) \\
F160W & 472 (78.5\%) &  313 (52.1\%) 
\enddata
\tablecomments{Valid measurements denote magnitudes that result from positive fluxes (signal measured above sky level) and suffer no other failures (e.g., image artifacts).  Well-determined measurements denote magnitudes where $\sigma < 0.5$ mag.}
\end{deluxetable*}


\begin{deluxetable*}{lcc}
\tablecaption{Luminosity Function Fits \label{lumfunfit}}
\tablehead{ \colhead{Passband} & \colhead{$\alpha_{L}$(All Clusters)} & \colhead{$\alpha_{L}$(Disk Clusters)}}
\tablewidth{0pt}
\startdata
F275W & $-1.72\pm0.07$ & $-1.76\pm0.08$ \\
F336W & $-1.73\pm0.06$ & $-1.90\pm0.08$ \\
F475W & $-1.79\pm0.06$ & $-2.06\pm0.09$ \\
F814W & $-1.74\pm0.05$ & $-2.18\pm0.09$ \\
F110W & $-1.68\pm0.06$ & $-2.20\pm0.11$ \\
F160W & $-1.56\pm0.05$ & $-1.82\pm0.08$ 
\enddata
\end{deluxetable*}


\begin{deluxetable*}{l|ccccr}
\tablecolumns{5}
\tablecaption{Background Galaxy Catalog \label{galcat}}
\tablewidth{0pt}

\tablehead{
\colhead{ID} & \colhead{Brick} & \colhead{RA} & \colhead{Dec} & \colhead{$R$}\\
\colhead{} & \colhead{} & \colhead{(J2000)} & \colhead{(J2000)} & \colhead{(arcsec)}
}

\startdata
  1 & 21 & 11.677181 & 42.192120 & 1.41\\
  2 & 21 & 11.644169 & 42.197941 & 1.20\\
  3 & 21 & 11.647937 & 42.202160 & 1.37\\
  4 & 21 & 11.685157 & 42.218735 & 0.97\\
  5 & 21 & 11.616282 & 42.196312 & 2.05
\enddata

\tablecomments{Table \ref{galcat} is published in its entirety in the electronic edition of the {\it Astrophysical Journal}.  A portion is shown here for guidance regarding its form and content.}
\end{deluxetable*}


\begin{deluxetable*}{lrlrlrlrlrlrlr}
\tabletypesize{\footnotesize}
\tablecaption{Summary of Existing Cluster Catalog Classifications and Revisions \label{oldcat-summary}}
\tablewidth{0pt}
\tablecolumns{14}

\tablehead{
\colhead{Catalog} & \multicolumn{2}{c}{Clusters} & \multicolumn{2}{c}{Candidates} & \multicolumn{2}{c}{Galaxies} & \multicolumn{2}{c}{\hii\ Regions} &
\multicolumn{2}{c}{Stars} & \multicolumn{2}{c}{Other\tablenotemark{a}} & \colhead{Total}
}

\startdata
RBC            & 78 & (71) & 4 & (34) & 3 & (3) & 1 & (3) & 98 & (76) & 3 & (0) & 187 \\
Kim (Total) & 12 & (10) & 2 & (28) & 2 & (0) & 0 & (0) & 22 & (0)    & 0 & (0) & 38 \\
Kim A          & 4 & (10) & 1 & (0) & 2 & (0) & 0 & (0) & 3   & (0) & 0 & (0) & 10 \\
Kim B          & 4 & (0) & 0 & (15) & 0 & (0) & 0 & (0) & 11 & (0) & 0 & (0) & 15 \\
Kim C          & 4 & (0) & 1 & (13) & 0 & (0) & 0 & (0) & 8   & (0) & 0 & (0) & 13 \\
Caldwell     & 68 & (73) & 1 & (0) & 0 & (0) & 1 & (0) & 113 & (110) & 0 & (0) & 183 \\
Peacock     & 77 & (58) & 4 & (25) & 2 & (2) & 0 & (2) & 41 & (40) & 3 & (0) & 127 \\
HKC            & 54 & (75) & 8 & (0) & 0 & (0) & 0 & (0) & 0 & (0) & 13 & (0) & 75 
\enddata

\tablecomments{In each column, the first number represents the number of objects per category after PHAT reclassification and the second number (in parenthesizes) represents the number of original classifications from the published catalog.  The A classification from \citet{Kim07} maps to a cluster classification in the RBC, while B and C map to candidate classifications.}
\tablenotetext{a}{The ``other'' classification signifies clusters that were not recovered by the PHAT search or were duplicate catalog entries in the RBC and \citet{Peacock10} catalogs.  In the case of the HKC, the ``other'' classification signifies objects that were deemed non-cluster asterisms.}
\end{deluxetable*}


\begin{deluxetable*}{lcccl}
\tablecaption{Revised Bologna Catalog Revisions \label{revise-rbc}}
\tablewidth{0pt}

\tablehead{
\colhead{Cluster Name} & \colhead{PC ID} & \colhead{New Flag} & \colhead{Old Flag} & \colhead{Comments}
}

\startdata
BH16       & 1381   & 1  & 2  & \nodata \\
B523       & 1383   & 1  & 2  & \nodata \\
SK118C     & 641    & 1  & 2  & \nodata \\
SK134C     & 1349   & 1  & 2  & \nodata \\
M028       & 544    & 1  & 2  & \nodata 
\enddata

\tablecomments{Table \ref{revise-rbc} is published in its entirety in the electronic edition of the {\it Astrophysical Journal}.  A portion is shown here for guidance regarding its form and content.}
\end{deluxetable*}


\begin{deluxetable*}{lcccl}
\tablecaption{Hodge-Krienke Catalog Revisions \label{revise-hodge}}
\tablewidth{0pt}

\tablehead{
\colhead{Cluster Name} & \colhead{PC ID} & \colhead{PHAT Classification}
}

\startdata
WH13	      & 1119   & Possible Cluster \\
WH18	      & 1089   & Possible Cluster \\
KHM31-195     & 1282   & Possible Cluster \\
KHM31-241     &  964   & Possible Cluster \\
Hodge09-57    &  977   & Possible Cluster 
\enddata

\tablecomments{Table \ref{revise-hodge} is published in its entirety in the electronic edition of the {\it Astrophysical Journal}.  A portion is shown here for guidance regarding its form and content.}
\end{deluxetable*}


\end{document}